\documentclass[fleqn,usenatbib]{mnras}

\usepackage{newtxtext,newtxmath}

\usepackage[T1]{fontenc}
\usepackage{ae,aecompl}

\usepackage{graphicx}	
\usepackage{amsmath}	
\usepackage{amssymb}	

\def\cmden {$\rm cm^{-3}$}
\def\cm2 {$\rm cm^{2} \,$}
\def\cm3 {$\rm cm^{3} \,$}
\def\kms {$\rm km\,s^{-1}$}
\def\ergcmspa {erg\,s$^{-1}$\,cm$^{-2}$\,spaxel$^{-1} \,$}

\def\ergs {erg\,s$^{-1}$}
\def\mjybeam {mJy\,beam$^{-1}$}
\def\msunyr {M$_\odot$\,yr$^{-1}$}
\def\o3 {[O\,{\sc iii}]\,}
\def\n2 {[N\,{\sc ii}]\,}

\newcommand{\etal}{et\thinspace al.\thinspace}

\title[Outflows in 4C\,+29.30] 
      {Powerful ionized gas outflows in the interacting radio galaxy 4C\,+29.30}

\author[Couto \etal]
       {Guilherme S. Couto$^1$\thanks{E-mail: guilherme.couto@uantof.cl},
        Thaisa Storchi-Bergmann$^2$,
        Aneta Siemiginowska$^3$,
        \and
        Rogemar A. Riffel$^4$,
        Raffaella Morganti$^{5,6}$
        \\
        $^{1}$Centro de Astronom\'ia (CITEVA), Universidad de Antofagasta, Avenida Angamos 601, Antofagasta, Chile \\
        $^{2}$Universidade Federal do Rio Grande do Sul, IF, CP 15051, Porto Alegre 91501-970, RS, Brazil \\
        $^{3}$Harvard Smithsonian Center for Astrophysics, 60 Garden St, Cambridge, MA 02138, USA \\
        $^{4}$Departamento de F\'isica, Universidade Federal de Santa Maria, Centro de Ci\^encias Naturais e Exatas, 97105-900, Santa Maria, RS, Brazil \\
        $^{5}$ASTRON, The Netherlands Institute for Radio Astronomy, Postbus 2, 7990 AA, Dwingeloo, The Netherlands \\
        $^{6}$Kapteyn Astronomical Institute, University of Groningen, PO Box 800, 9700 AV, Groningen, The Netherlands \\
       }

\date{Accepted XXX. Received YYY; in original form ZZZ}

\pubyear{2020}

\begin{document}
\label{firstpage}
\pagerange{\pageref{firstpage}--\pageref{lastpage}}
\maketitle

\begin{abstract}

We investigate the ionized gas excitation and kinematics in the inner $4.3\,\times\,6.2$\,kpc$^{2}$\, of the merger radio galaxy 4C\,+29.30. Using optical integral field spectroscopy with the Gemini North Telescope, we present flux distributions, line-ratio maps, peak velocities and velocity dispersion maps as well as channel maps with a spatial resolution of $\approx 955\,$pc. We observe high blueshifts of up to $\sim-650\,$\kms\, in a region $\sim1''$ south of the nucleus (the southern knot -- SK), which also presents high velocity dispersions ($\sim 250\,$\kms), which we attribute to an outflow. A possible redshifted counterpart is observed north from the nucleus (the northern knot -- NK). We propose that these regions correspond to a bipolar outflow possibly due to the interaction of the radio jet with the ambient gas. We estimate a total ionized gas mass outflow rate of $\dot{M}_{out} = 25.4 \substack{+11.5 \\ -7.5}\,$\msunyr with a kinetic power of $\dot{E} = 8.1 \substack{+10.7 \\ -4.0} \times 10^{42}\,$\ergs, which represents $5.8 \substack{+7.6 \\ -2.9} \%$ of the AGN bolometric luminosity. These values are higher than usually observed in nearby active galaxies with the same bolometric luminosities and could imply a significant impact of the outflows in the evolution of the host galaxy. The excitation is higher in the NK -- that correlates with extended X-ray emission, indicating the presence of hotter gas -- than in the SK, supporting a scenario in which an obscuring dust lane is blocking part of the AGN radiation to reach the southern region of the galaxy.

\end{abstract}




\section{Introduction}

Active Galactic Nuclei (AGN) feedback is now believed to play a major role in galaxy evolution. In order to explain the observed scaling relationships between the supermassive black holes (SMBHs) and galaxy bulge properties, such as the mass and stellar velocity dispersion \citep[$M_\bullet$\,--\,$\sigma_\ast$,][]{ferrarese00,gebhardt00,mcconnell13,kormendy13}, AGN feedback is usually summoned. It is indeed required in evolutionary simulations to reproduce star formation quenching in early type galaxies, preventing them from becoming too massive \citep{fabian12,heckman14,bongiorno16,su19}. In this scenario, radio galaxies are particularly interesting to study the effect of outflows in AGN hosting galaxies. Radio loud AGNs in particular present jets that may reach distances beyond the optical extent of the galaxies \citep{bridle84}. These jets expand through the galaxy interstellar medium (ISM), affecting the energetics and thermodynamics of the gas, and possibly generating gas ionization by shocks \citep{groves04b,wagner11}. When resolved, emitting gas regions, which often appear aligned with the radio jet \citep{miley81,cecil88,tremblay09,dasyra15,mahony16}, are important tracers of AGN feedback, allowing for a detailed study of their kinematics and their role in the AGN energetic budget.

Although AGN feedback processes -- and in particular gas outflows in radio galaxies -- have been identified in recent works \citep[e.g.,][]{santoro15,couto17,revalski18,venturi18}, relations between AGN and its outflows have just recently begin to be explored \citep[e.g.][]{fiore17} in order to determine whether or not outflows deliver effective power, capable of altering star formation rates and evacuating gas reservoirs \citep{harrison17,zubovas17}. AGN feedback quantification can be accomplished by characterizing the outflow kinetic energy, mass outflow rate, and related energetic measurements. Determining these quantities is greatly aided by spatially resolved observations that constrain the physical size and location of the outflowing material, making nearby AGN interesting targets as observations can resolve outflows on sub-kiloparsec scales. In this sense, Integral Field Spectroscopy (IFS) observations are key, since they provide a direct observation of the impact of the AGN on the ambient gas, allowing to measure the gaseous kinematics and ionization and its relation with the feedback.

In previous studies investigating the impact of radio jets in the evolution of the host galaxies, it has been concluded that it seems to be high mostly in galaxy clusters environments \citep{fabian12,russel19} while inside the galaxy the impact seems to be modest \citep{couto13,couto17}. This latter result may be due to the fact that we have ``missed the action" in our previous targets and have not looked at radio galaxies in which strong feedback -- e.g. via interaction of the radio jets with the ambient gas -- was still occurring. In fact, indications have been found that young or newly restarted jets are the one where this interaction is stronger \citep{holt08,holt09,rodriguez-ardila17}. In order to verify the local feedback (inside the galaxy) scenario, we have obtained IFS observations of the inner few kpc of 4C\,+29.30, a radio galaxy with an early-type morphology host at redshift $z = 0.0647$ and moderate radio luminosity ($\sim 10^{42}\,$\ergs), which presents signatures of jet reactivation. The corresponding luminosity distance is 289\,Mpc and 1 arcsec corresponds to $1.24\,$kpc in a cosmology with $H_0 = 70.5\,$\kms\,Mpc$^{-1}$, $\Omega_\Lambda = 0.73$ and $\Omega_M = 0.27$. 

4C\,+29.30 is particularly interesting because multiple episodes of activity have been revealed from the morphology and spectral properties of the radio emission over a broad range of scales. It was first studied in the radio and optical bands by \citet{vanbreugel86}, who found optical line emitting gas extending to $\sim 20''$ north of nucleus, and adjacent to the radio jet along a position angle P.A. $= 24^\circ$, an evidence of the radio jet interacting with dense extranuclear gas. 4C\,+29.30 is possibly a merger system, displaying a characteristic dust lane passing in front of the central region in similar fashion to Centaurus A. A low surface-brightness, radio diffuse emission extended to $\sim600\,$kpc has been detected and studied by \citet{jamrozy07}. This structure is characterised by a very steep radio spectrum typical of radio emission from remnants activity, i.e. not fuelled anymore by the central core. Based on the spectral properties, the authors derived the age of this structure to be $\ge 200\,$Myr. On the $\sim 40\,$kpc scale, a double-lobed source resulting from on-going activity with estimated age $\le 30\,$Myr.   

The central region, i.e. the inner $\sim 20\,$pc, have been imaged by \citet{liuzzo09} using the VLBA and VLBI networks. Interestingly, a double-lobed source elongated in the same direction as the kpc-scale structure, is observed on these scales. Although the spectral information is not available at such high spatial (milliarcsec) resolution, the authors noted that the structure of the source suggests it is the result of a recently restarted phase of activity. They estimated this restarted phase to have an age of $\sim 10^4\,$yr. Thus, the three structures visible in the radio morphology on very different scales indicate at least three phases of recurrent radio activity, similar to other radio galaxies such as Centarus A, B2\,0258+35, J1216+0709 and many others \citep{kuzmicz17,mckinley18,morganti99,brienza18,singh16}. On X-rays bands, data show a complex view of interactions between jet-driven radio outflows and the host galaxy environment, signaling feedback processes closely associated with the central absorbed active nucleus \citep{siemiginowska12,sobolewska2012}.


Our study consists of a two-dimensional analysis of the gas excitation and kinematics of the inner $4.3 \times 6.2\,$kpc$^2$ of 4C\,+29.30 and is organized as follows. In Section \ref{obs} we describe the observations and data processing; in Section \ref{res} we present our strategy to obtain emission-line excitation and kinematic constraints, as well as the results of our measurements; we discuss our results and present possible scenarios to explain our observations in Section \ref{dis}; finally we present our main conclusions in Section \ref{conc}.

\section{Observations and Data Reduction}
\label{obs}

\subsection{Gemini GMOS data}

4C\,+29.30 was observed between January 12 and March 15, 2016, with the Integral Field Unit (IFU) of the Gemini Multi-object Spectrograph \citep[GMOS;][]{allington02} mounted on the Gemini North Telescope (Gemini program ID GN-2016A-Q-77). ``One-slit'' mode was used, with a rectangular field-of-view (FoV) of $\approx 3\farcs5 \times 5\farcs0$, corresponding to $4.3\,\times \, 6.2\,$kpc$^{2}$ at the galaxy. Fifteen exposures of 1140\,s were obtained, slightly shifted and dithered (up to 0\farcs8 in both axes) in order to correct for detector defects after combination of the frames.

Spectra with wavelength coverage in the range $\lambda4500 - 7300\,$\AA, centered at $\lambda5900\,$\AA, were obtained with the use of the B600+\_G5307 grating and IFU-R mask. The spectral resolution is $R \sim 3600$ at $\sim \lambda 7000\,$\AA\, ($\sim83$\,\kms), derived from the full width half maximum (FWHM) of the CuAr emission lines. Spectral dithering was also performed, with a maximum separation of $102.5$\AA\, between exposures.

The data reduction was performed using the {\sc IRAF} \citep{tody86,tody93} packages provided by the Gemini Observatory, and specifically developed for the GMOS instrument. The procedure consists of sky and bias subtraction, flat-fielding, trimming, wavelength and relative flux calibration, building of the datacubes, final alignment and average combination with an average sigma clipping into the final datacube, which has a spatial binning of $0.1 \times 0.1\,$arcsec$^{2}$. 

The program did not include standard stars observations, and we used Feige 66, taken from the Gemini archive, and a Sloan Digital Sky Survey (SDSS) $3''$ aperture integrated spectrum of 4C\,+29.30, to provide a relative calibration of the resulting galaxy datacube fluxes. Using star profiles in the acquisition images of each observation night, we measured a mean FWHM of $\approx 0\farcs77$, and we consider this to be  the angular resolution of the final datacube. This corresponds to $\approx 955\,$pc at the galaxy.

\subsection{Ancilliary data}

In this work we make use of archival data from previous studies of 4C\,+29.30 to compare with and help in the analysis of our IFS data. Here we briefly describe these data. 

\subsubsection{{\it HST}-STIS}

The {\it Hubble Space Telescope}, using the Space Telescope Imaging Spectrograph (STIS), observed 4C\,+29.30 in January 25, 2001, with the MIRVIS filter. Broad-band imaging was centered at $\lambda 5852\,$\AA\, with a bandwidth of $\lambda 1873\,$\AA\, and a total exposure time of 2683\,s. 

\subsubsection{VLA}

4C\,+29.30 was observed by the Very Large Array (VLA) in the frequency of 4.8 GHz in the A configuration in February 26, 1982, with a $0\farcs3\, \times\, 0\farcs3$ resolution. Further information on the data can be found in \citet{vanbreugel86}. 

\subsubsection{{\it Chandra}}

Deep {\it Chandra} Advanced CCD Imaging Spectrometer (ACIS) imaging observations of 4C\,+29.30 were performed in February 2010, in energies between 0.5\, keV and 7\,keV (soft and hard bands), with a total exposure of 286.4 ks. See \citet{siemiginowska12} for a detailed description of the data.

\section{Results}
\label{res}

\begin{figure*}
\centering
\includegraphics[width=\textwidth]{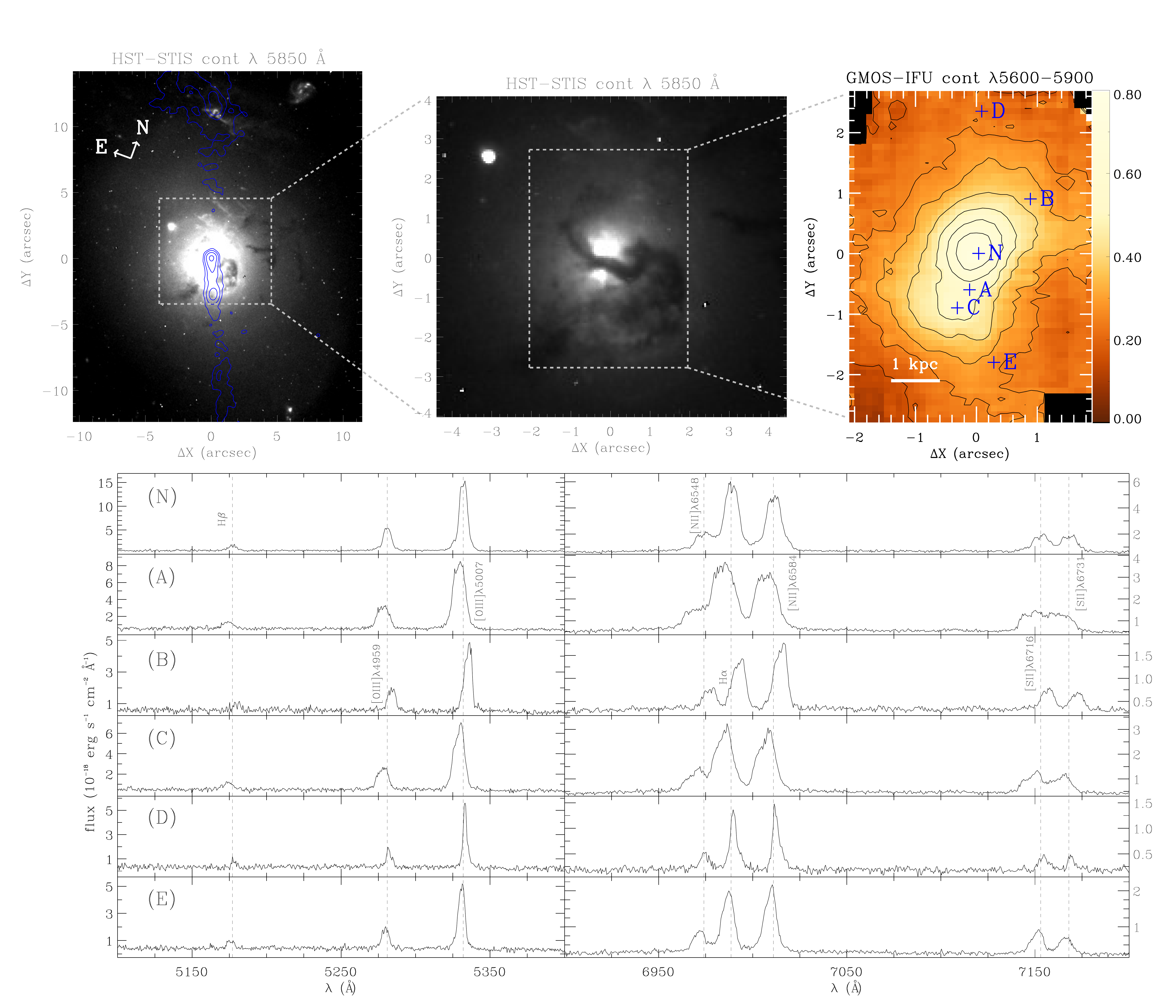}
\caption{Top left and central panels: {\it HST}-STIS continuum optical image of 4C\,+29.30. Top right panel: GMOS-IFU continuum image. Bottom panels: Spectra extracted in positions N, A, B, C, D and E, which are identified in the top right panel. Blue contours display VLA 4.8 GHz radio emission. Flux units of the continuum map are $10^{-18}\,$\ergcmspa. Radio contour levels are 0.1, 0.4, 1.6 and 6.4 \mjybeam. Note that all maps are oriented as indicated by the arrows in the top left panel, with the north making an angle of $20^\circ$ clockwise with the vertical axis of the frames (northeast is up and southeast to the left).}
\label{large}
\end{figure*}

In Fig. \ref{large} we display a general look of the galaxy via {\it HST} continuum image, showing also the IFU field-of-view and the corresponding IFU data and structures of 4C\,+29.30 on different scales. The top left and central panels display the same {\it HST}-STIS optical continuum image in two different scales and count limits. The top left image has a FoV of $22\,\times\,26\,$arcsec$^2$ and shows many dust clouds and filaments, as well as what seems to be bright gas emitting regions reaching up to $\sim 12''$ north of the nucleus. A dust lane crosses the region just below the nucleus as seen to the top central panel, which is a closer look into the inner $\sim 8''\,\times\,8''$ region of the left image of 4C\,+29.30. Blue contours trace the VLA 4.8 GHz radio image overploted in the top left image extending to the north-east and south-west. In the top right panel we show a continuum image of 4C\,+29.30, obtained using our GMOS-IFU datacube (FoV of $\sim 3\farcs5\,\times\,5''$), extracted between wavelengths $\lambda5600-5900\,$\AA. Due to the lower angular resolution compared to {\it HST} data, instead of the abrupt transition due to the dust lane, we observe a smooth transition between the top knot of emission (whose center we identify with the galaxy nucleus and will be identified by a ``+'' sign in the maps shown in this paper) and bottom emission knot that we from now on call ``the southern knot'' (hereafter SK, where position C is located). The three top panels have the same orientation, with the $y$-axis along P.A.\,$= 20^\circ$, very close to the orientation of the jet axis. 

In the bottom panels we display spectra extracted in the marked positions N (nucleus), A ($\approx 0\farcs6$ south from the nucleus), B ($\approx 1\farcs3$ north-west from the nucleus), C ($\approx 0\farcs9$ south from the nucleus), D ($\approx 2\farcs4$ north-east from the nucleus) and E ($\approx 1\farcs8$ south-west from the nucleus). Each spectrum corresponds to one spaxel, thus $0\farcs1 \times 0\farcs1$. We display the spectral regions showing the strongest emission lines: H$\beta$, [O\,{\sc iii}]$\lambda 4959,5007$, [N\,{\sc ii}]$\lambda 6548,84$, H$\alpha$ and [S\,{\sc ii}]$\lambda 6716,31$. The emission lines present complex profiles suggesting the presence of more than one kinematic component, mainly in spectra A, B and C. Redshifted emission lines are seen in the north-western part of the FoV, as observed in B and D spectra, while A, C and E, located in the south-eastern part of the FoV, present mainly blueshifted velocities. Large line broadening is observed in the spectrum A, and somewhat smaller in spectra N, B and C, while narrow profiles are observed only in the border of the FoV, as seen in spectrum D. The dashed lines display the emission lines wavelengths at the systemic velocity of the galaxy.

\subsection{Emission-line measurements}

Gauss-Hermite profiles were fitted to the most prominent emission lines (H$\beta$, [O\,{\sc iii}]$\lambda\lambda 4959,5007$, [O\,{\sc i}]$\lambda\lambda 6300,34$, [N\,{\sc ii}]$\lambda\lambda 6548,84$, H$\alpha$ and [S\,{\sc ii}]$\lambda\lambda 6716,31$) in order to derive peak velocities, velocity dispersions and integrated fluxes. We used customized IDL\footnote{IDL, or Interactive Data Language, is a programming language used for data analysis and visualization.} routines with the \textsc{mpfitfun} module for the measurements. Gauss-Hermite polynomials were chosen to take into account the asymmetries in the emission line profiles, and are expressed by the following equations:

\begin{equation}
f_{gh} (\lambda) = F {\frac{e^{-k^2/2}}{\sqrt{2\sigma^2}}} (1+h_3H_3+h_4H_4)
\end{equation}

\begin{equation}
H_3 = {\frac{1}{\sigma}} (2\sqrt{2}k^3 - 3\sqrt{2}k)
\end{equation}

\begin{equation}
H_4 = {\frac{1}{\sqrt{24}}} (4k^4 - 12k^2 + 3)
\end{equation}

\begin{equation}
k = {\frac{(\lambda - \overline{\lambda})}{\sigma}}
\end{equation}

\noindent
where $F$ is the flux, $\sigma$ is the velocity dispersion and $\overline{\lambda}$ is the peak wavelength. 

The $h_3$ and $h_4$ Gauss-Hermite moments parameterize the deviations from a Gaussian profile, thus are good tracers of multiple emission-line components. $h_3$ is related to the skewness of the profiles, and $h_4$ to its kurtosis. In other words, $h_3$ measures asymmetric deviations from a Gaussian profile, such as blue ($h_3<0$) or red ($h_3>0$) wings, and $h_4$ quantifies the peakiness of the profile, with $h_4 > 0$ for a more peaked and $h_4 < 0$ for a broader profile than a Gaussian curve. A Gaussian profile is obtained when $h_3 = h_4 = 0$. 

To reduce the number of free parameters in the fit, the following physically motivated constraints
were imposed:

\begin{itemize}
\item different lines from the same ionic species have the same kinematic parameters. For example, the [S\,{\sc ii}]$\lambda 6716,31$ emission lines have the same peak velocity and velocity dispersion. This was also done for the Gauss-Hermite parameters $h_3$ and $h_4$;
\item the [N\,{\sc ii}]$\lambda\lambda 6548,84$ emission lines have the same peak velocity and velocity dispersion as H$\alpha$;
\item the [N\,{\sc ii}]$\lambda 6548$ flux was fixed as 1/3 of the [N\,{\sc ii}]$\lambda 6584$ flux, in accordance with nebular physics \citep{osterbrock06}. This was also done for the [O\,{\sc iii}]$\lambda\lambda 4959,5007$ and [O\,{\sc i}]$\lambda\lambda 6300,34$ emission lines.
\end{itemize}

In order to estimate the errors on the measured quantities we performed Monte Carlo simulations: for each spaxel, we constructed one hundred realizations of the spectrum by adding Gaussian noise with amplitude comparable to the noise measured in the original spectrum. Mean values and standard deviations for the peak velocities, velocity dispersions and fluxes were derived for each spaxel, with the standard deviation of the distribution in each parameter being adopted as the uncertainty.

\begin{figure}
\centering
\includegraphics[width=0.5\textwidth]{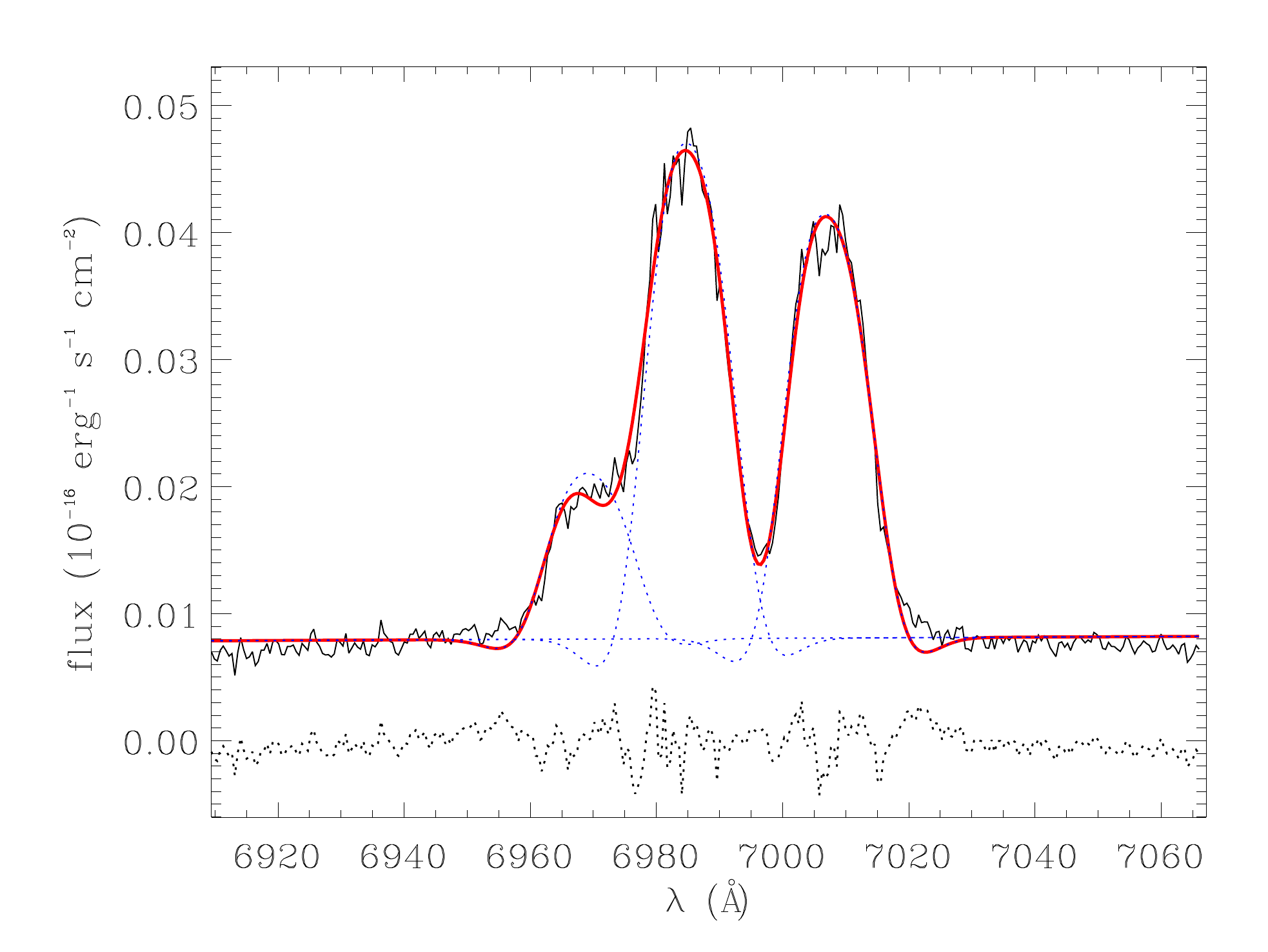}
\caption{Fit of Gauss-Hermite polynomials to the
[N\,{\sc ii}]+H$\alpha$ emission-line profiles, for position A from top right panel of Fig. \ref{large}. Spectrum data is shown by the black line, while the red line represent the best model fitted to the profiles, with the components of each emission line shown by dotted blue lines. Residuals are shown by dotted black lines.}
\label{gh_fit}
\end{figure}

Fig. \ref{gh_fit} shows an example of a fit of the [N\,{\sc ii}]+H$\alpha$ emission lines using the Gauss-Hermite profiles. This spectrum is in the position identified as `A' in Fig. \ref{large}, where a very broad profile is observed, as well as a broad peak, and since $h_4$ is negative in this profile. As shown, residuals are small and in the level of the noise.

\subsection{Emission-line flux, peak velocity and velocity dispersion distributions}

\begin{figure*}
\centering
\includegraphics[width=\textwidth]{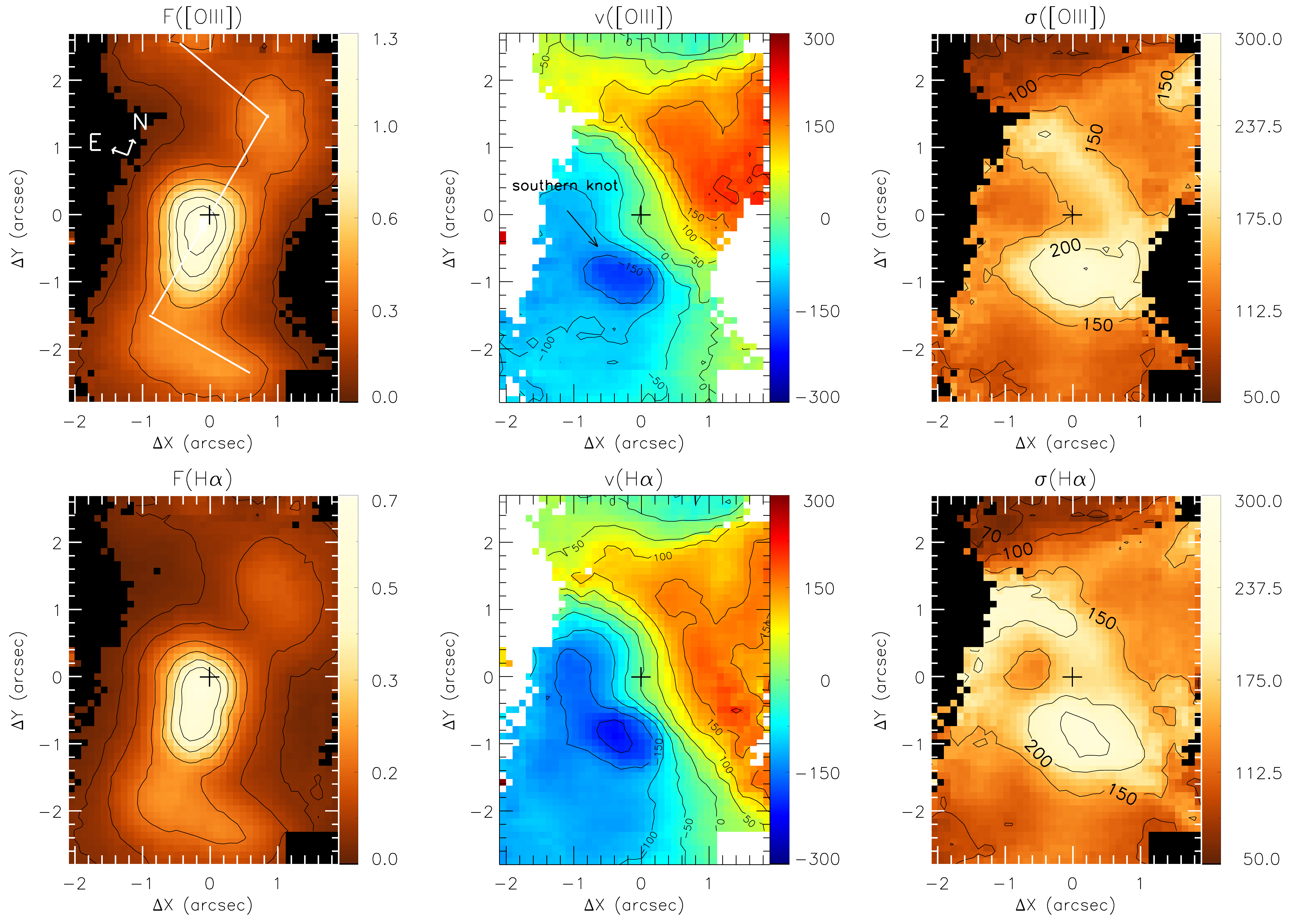}
\caption{Integrated flux (left panels), peak velocity (central panels) and velocity dispersion (right panels) distributions resulted from the Gauss-Hermite fit of the [O\,{\sc iii}]$\lambda 5007$ (top panels) and H$\alpha$ (bottom panels) emission lines. The white lines in the top left panel indicate the ``z-shaped'' structure mentioned in the text. The southern knot location is also indicated in the [O\,{\sc iii}] peak velocity map. Flux units are $10^{-16}\,$\ergcmspa. Peak velocity and velocity dispersion units are \kms.}
\label{flux_velsig}
\end{figure*}

Fig. \ref{flux_velsig} displays the integrated line flux (not extinction corrected), peak velocity and velocity dispersion distributions, for the H$\alpha$ and [O\,{\sc iii}]$\lambda 5007$ emission lines. The emission-line flux maps are somewhat different from the continuum flux map we have shown in the top right panel of Fig. \ref{large}. Gas emission appears extended along the south-north direction, with the emission to the south bending to the west and that to the north bending to the east close to borders of the FoV ($\gtrsim 2\farcs0$) in a ``z-shaped" structure. Close to the galaxy nucleus (identified with  the peak of the continuum emission), we observe stronger emission to the south than to the north, which extends from the nucleus to $\approx$\,1$^{\prime\prime}$ south of it. The northern extended emission appears to be ``clumpier'' than the southern emission, with peak at $\approx 1\farcs8$ from the nucleus. The dust lane can be identified as the decrease in brightness in the continuum image (shown in Fig. \ref{large}); it is not as sharp in our flux maps as in the HST images due to the poorer angular resolution of the GMOS-IFU data, but we do observe some decrease in the continuum emission just to the south of the nucleus along the east-west direction that can be attributed to the dust lane.

A distorted rotation pattern is observed in the peak velocity maps (central panels of Fig. \ref{flux_velsig}), with blueshifts to the south and south-east of the nucleus and redshifts to the north and north-west. The redshifted region appears to show less kinematic structure than the blueshifted region, with a velocity amplitude of $\sim 170\,$\kms. In the blueshifted region we observe a steeper increase of blueshifted velocities (amplitudes of $\sim -200\,$\kms) that we identify with the SK described in the previous section. Another kinematic structure is observed beyond $\approx 2''$ north-east from the nucleus, with velocities decreasing to zero towards the top border of our FoV, giving the impression that blueshifts would be observed further out if probed by our FoV. We have estimated a systemic velocity of $v_{\mathrm{sys}} = 19487.9\,$\kms\,as the mean velocity in a $5\times5$ pixels region centered in the nucleus, considering both [O\,{\sc iii}] and H$\alpha$ velocity fields. This value is comparable with other estimates in the literature \citep[e.g. $v = 19439.0\pm5\,$\kms,][]{lavaux11}, considering our spectral resolution. The systemic velocity was subtracted from the original velocity maps in order to obtain the peak velocity fields shown in Fig. \ref{flux_velsig}.

The highest velocity dispersions ($\sim 250\,$\kms) are observed approximately co-spatial to the region presenting the highest blueshifted velocities, although somewhat ($\approx 0\farcs5$) shifted west from it. In fact, this region is close to the position A which presents broad emission-line peaks (see the corresponding spectrum in Fig. \ref{large}). Somewhat lower values of $\sigma \approx 170\,$\kms\, (although larger than values from the surroundings) are observed in a $\approx 1\farcs0$ wide strip just to the north of the nucleus, crossing the region from north-east to south-west. This region corresponds to a location in which the peak isovelocity contours are very close to each other, showing a steep variation from $\sim -50$ \kms\, to $\sim 100$ \kms. Velocity dispersion values then drop at farther regions, reaching $\sigma \approx 100\,$\kms\, by $\approx\,2''$ from the nucleus, with  the smallest values of $\sigma \approx 70\,$\kms\, being observed in the north-east border of the FoV, at the region where zero velocities are observed in the peak velocity maps.

Expressing the flux uncertainty, $\epsilon_F$, as a fraction of the integrated flux, $F$, we find that H$\alpha$ and [O\,{\sc iii}] flux maps show typical values $\epsilon_F / F \approx 0.01$ in the nucleus and along the extended emission to the north and south. $\epsilon_F / F \approx 0.1 - 0.2$ are observed toward the east and west, and the border of the FoV in these directions present $\epsilon_F / F \approx 0.3$. The uncertainties of the peak velocity and velocity dispersion maps are similar and present values of $\epsilon_v \approx \epsilon_\sigma \approx 5\,$\kms\, in the nucleus and the regions of strongest extended emission, with values of $20\,$\kms\, closer to the borders of the FoV. We have masked out, in Fig. \ref{flux_velsig} and in the other maps presented in this paper, regions that present uncertainty values higher than $\epsilon_F / F = 0.2$, $\epsilon_v = 20\,$\kms\, or $\epsilon_\sigma = 20\,$\kms, for each emission line.

\subsubsection{$h_3$ and $h_4$ Gauss-Hermite parameters distributions}

\begin{figure*}
\centering
\includegraphics[width=0.7\textwidth]{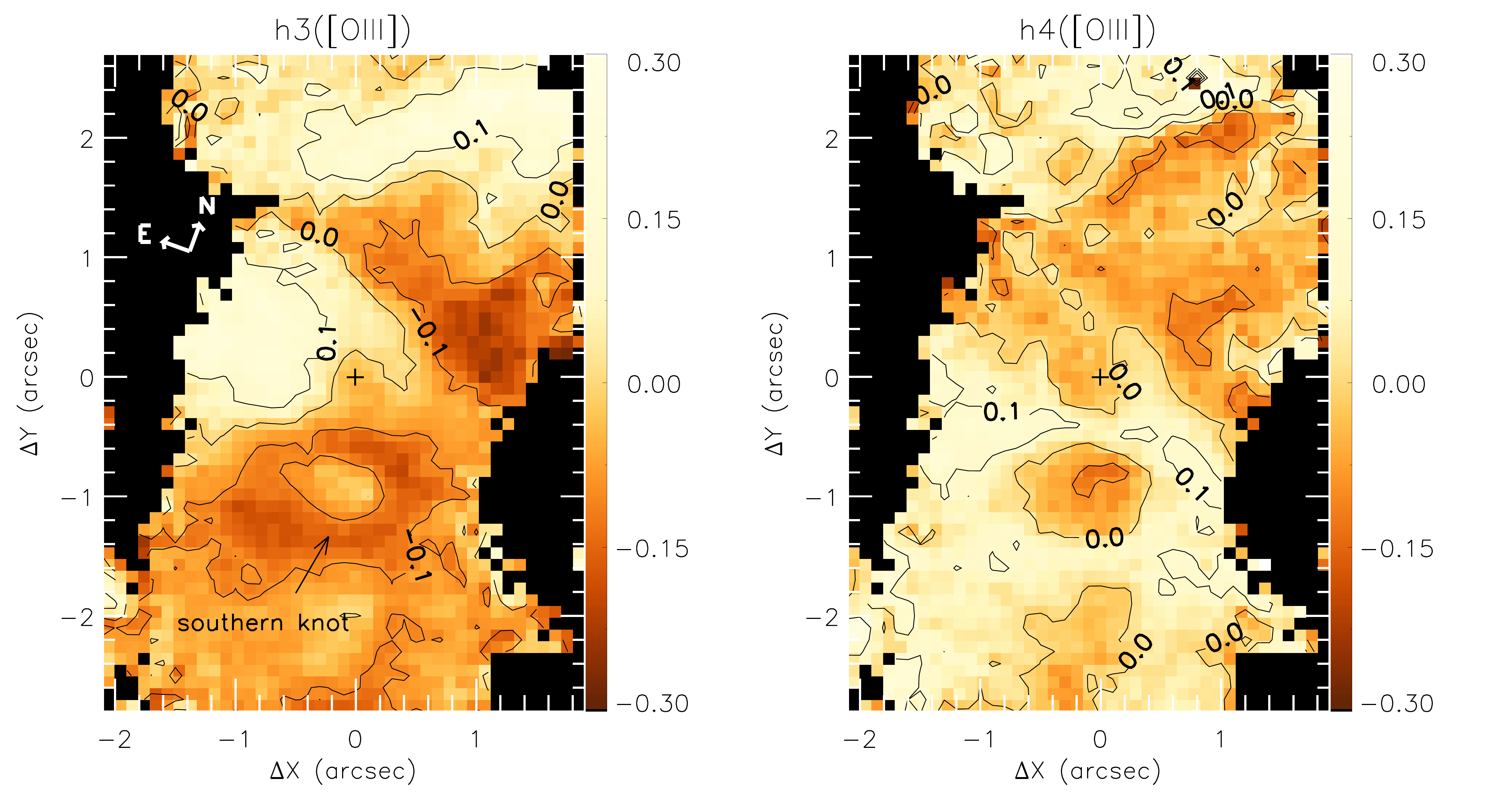}
\caption{Distribution maps of Gauss-Hermite $h_3$ and $h_4$ moments, obtained from the fit of the [O\,{\sc iii}]$\lambda\lambda 4959,5007$ emission lines. These moments indicate asymmetry ($h_3$) and peakiness/broadening ($h_4$) of the line profiles. The southern knot location is indicated in the $h_3$ map.}
\label{h_3_h_4}
\end{figure*}

The distribution maps of the Gauss-Hermite moments characterizing an asymmetry and peakiness of line profiles are shown in Fig. \ref{h_3_h_4}. The values of the two moments vary throughout our FoV, indicating more than one kinematical component and complex variations. The region presenting redshifted peak velocities, located $\approx 1\farcs4$ north-north-west from the nucleus, shows $h_3$ values smaller than $-0.1$, indicating the presence of blue wings and negative $h_4$ values, indicating broader profiles. The opposite is observed in the region showing blueshifts $\approx 0\farcs9$ east from the nucleus, where the $h_3$ moment presents values higher than $0.1$, tracing red wings. The blueshifted SK ($\approx 1''$ from the nucleus), on the other hand, presents a ring of low $h_3 \sim -0.1$ and $h_4 \sim 0$, surrounding a region with $h_3 \sim 0$ and $h_4 \sim -0.1$, showing that this region is dominated by a characteristic kinematic component apart from the ones observed around it, with a broad peak in the emission lines profiles. The zero (to slightly blueshifted) velocity region at the north-east border of the FoV shows $h_3 \sim 0.1$, thus red wings. The $h_4$ moment distribution indicates a more peaked profile in the blueshifted regions south to south-east from the nucleus (except for the SK), while a broader profile is present in the redshifted regions north-west from the nucleus and in the SK.

\subsection{Channel maps}

\begin{figure*}
\centering
\includegraphics[width=\textwidth]{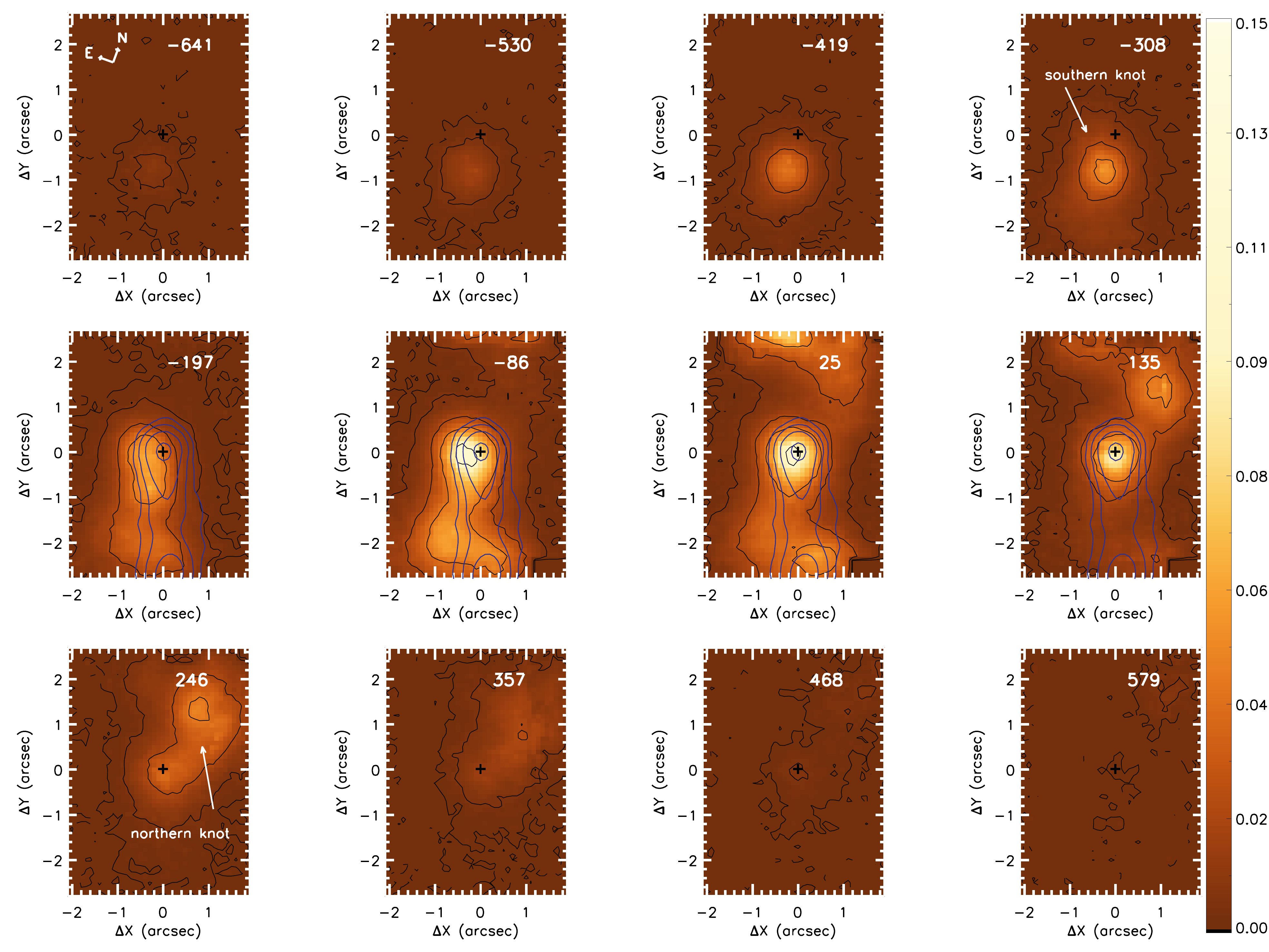}
\caption{Channel maps along the [O\,{\sc iii}]$\lambda 5007$ emission-line profile, in order of increasing velocities shown at the top of each panel in units of \kms. Flux units are $10^{-16}\,$\ergcmspa. The blue contours display the radio jet. Radio contour levels are 0.1, 0.4, 1.6 and 6.4 \mjybeam.}
\label{cmOIII}
\end{figure*}

We have mapped the gas kinematics using also channel maps extracted along the [O\,{\sc iii}]$\lambda 5700$ emission line profile. Fig. \ref{cmOIII} shows a sequence of channel maps within velocity bins of $\approx 111\,$\kms\, (corresponding to four spectral pixels). The highest blueshifts ($\sim -650\,$\kms) are observed in the SK, whose emission is observed  down to blueshifts $\sim -200\,$\kms; for less negative velocities, a linear structure extends to the south of the nucleus and for velocities closer to zero, another knot seems to appear at the bottom (south-west) border of the FoV, and a partial knot can also be seen at the top (north-east) border of the FoV. The highest emission is seen at the nucleus for velocities in the range $-197 < v < 135\,$\kms, along with the southern structure. For higher redshifts ($135 < v < 357\,$\kms) a northern knot (hereafter NK) of emission is observed $\approx\,1\farcs8$ from the nucleus, whose emission, together with that of the nucleus, can be observed up to 468\,\kms. We have identified both the SK and the NK in Fig. \ref{cmOIII}.

The radio jet, represented by the blue contours, show spatial correlation with the nucleus and the knot of emission at the bottom border of the map, $\sim 2''$ from the nucleus, where a increase of the radio emission is observed. This is clearer in the channel map with $v \sim 25\,$\kms.

\subsection{Line ratio distributions}

\begin{figure*}
\centering
\includegraphics[width=\textwidth]{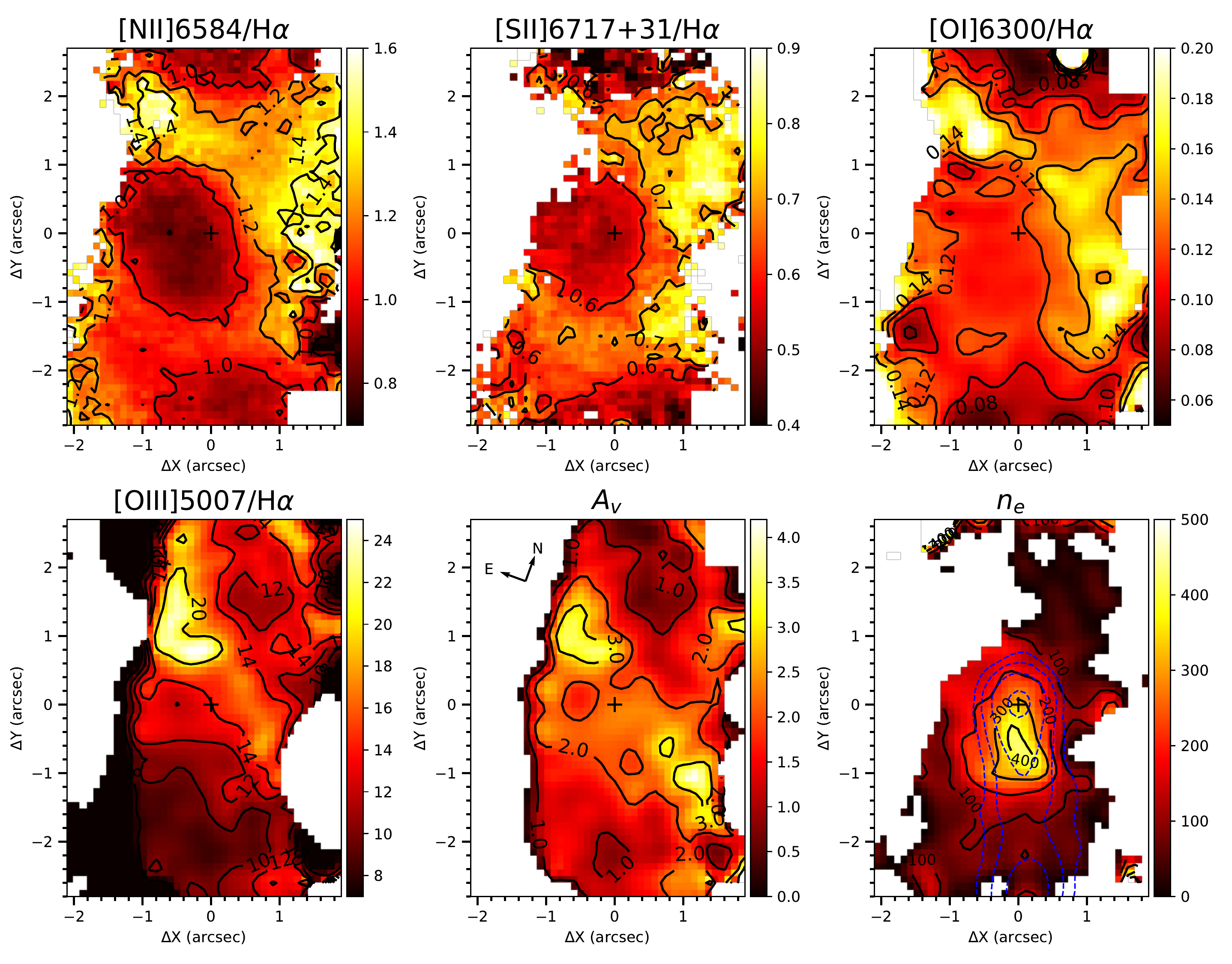}
\caption{Emission-line ratio maps (top and bottom left panels), visual extinction (bottom center, units are in magnitudes) and electron density (bottom right, units are \cmden). The dashed blue contours on the density map display the VLA 4.8 GHz radio emission, and contour levels are 0.1, 0.4, 1.6 and 6.4 \mjybeam.}
\label{ratios}
\end{figure*}

The top left and middle panels of Fig. \ref{ratios} display the [N\,{\sc ii}]/H$\alpha$ and [S\,{\sc ii}]/H$\alpha$ ratio maps, which are similar to each other, and seem to be spatially correlated with the peak velocity maps of Fig. \ref{flux_velsig}: the highest ratio values ([N\,{\sc ii}]/H$\alpha \sim 1.5$ and [S\,{\sc ii}]/H$\alpha \sim 0.85$) are observed in the redshifted region north-west of the nucleus, while the lowest ratio values ([N\,{\sc ii}]/H$\alpha \sim 0.8$ and [S\,{\sc ii}]/H$\alpha \sim 0.5$) are observed in the blueshifted regions, east and south-east of the nucleus (and including it). The approximately zero velocity region at the north-eastern border of the FoV also shows a decrease of these line ratio values. The blueshifted SK displays no particular structure in the line-ratio maps. The [O\,{\sc i}]/H$\alpha$ ratio is low throughout our FoV, with the highest values of up to $\sim 0.2$, in the regions of high [N\,{\sc ii}]/H$\alpha$ and [S\,{\sc ii}]/H$\alpha$ ratios, where redshifts are observed in the peak velocity maps.

High [O\,{\sc iii}]/H$\beta$ ratio values, of $\sim 15$, are observed at the nucleus and towards the east to north-east, reaching its highest value ($\sim 25$) about $1''$ from it. The lowest ratios ($\sim 9$) are observed $\approx\,2''$ south from the nucleus.

The visual extinction $A_V$ was obtained from the H$\alpha$/H$\beta$ ratio, and its map is shown in the bottom central panel of Fig. \ref{ratios}. We adopted the reddening law from \citet{cardelli89} and assumed case B recombination from \citet{osterbrock06}, leading to:

\begin{equation}
A_V = R_V\,E(B-V) = 6.9 \times log \left( {\frac{H\alpha/H\beta}{3.1}} \right) \, .
\end{equation} 

The dust lane is clearly the main feature, along which $A_V$ is at least 2.0 and reaches $A_V > 3.0$ at some locations. Then $A_V$ decreases both to the north and south of the nucleus. The relatively high values of $A_V > 1.0$ over most of the FoV indicates a high concentration of dust in the inner $\sim 2.5\,$kpc radius of 4C\,+29.30.

In order to calculate the emitting gas density, we used the ratio of the [S\,{\sc ii}] emission lines \citep[bottom right panel of Fig. \ref{ratios},][]{osterbrock06} and the {\sc PyNeb} routine \citep{luridiana15}, for a typical temperature of $10\,000\,$K. The electron density map is shown in the bottom right panel of Fig. \ref{ratios}. The highest densities reach values greater than $400\,$\cmden\, at the nucleus and extending $\sim 1\farcs0$ to the south-south-west, covering the dust lane and the SK. The radio contours seem to be co-spatial with this region of highest density. The electron density then decreases outwards, with the lowest values of $\sim 50\,$\cmden\, observed to the north of the nucleus at the locations with enhanced line ratios of [N\,{\sc ii}]/H$\alpha$, [S\,{\sc ii}]/H$\alpha$ and [O\,{\sc i}]/H$\alpha$.

The [O\,{\sc i}]$\lambda 6300$/H$\alpha$, [O\,{\sc iii}]/H$\beta$ and H$\alpha$/H$\beta$ (and thus the visual extinction) ratio maps were constructed using smoothed flux maps of its corresponded emission-lines. We applied a spatial Gaussian filter with a $0\farcs3$ radius to the original flux maps, which is smaller than our seeing-limited spatial resolution, in order to improve the signal to noise ratio of H$\beta$ and [O\,{\sc i}]$\lambda 6300$ lines, which are very weak.
 
\section{Discussion}
\label{dis}

\subsection{Excitation}

\subsubsection{Flux distributions}

4C\,+29.30 clearly shows evidence of interaction, as supported by the presence of dust lanes crossing the nuclear region of the galaxy seen in the HST image (Fig.\,\ref{large}). The [O\,{\sc iii}] and H$\alpha$ flux distribution maps display the maximum emission co-spatial with the peak of the continuum emission -- assumed to be the nucleus of the galaxy, with an elongation at high emission levels to $\sim 1''$ south from the nucleus. This elongation is the combination of the nuclear flux and that of the SK, which are separated in the {\it HST} image by the dust lane seen in the {\it HST} image of Fig.\,\ref{large}, but that becomes smoothed by the poorer angular resolution of the GMOS-IFU data ($\sim 0\farcs77$). Another knot of emission is observed $\approx 1\farcs8$ north from the nucleus. While the SK can be identified with the blueshifted knot seen in the gas velocity field (Fig.\,\ref{flux_velsig}) and negative velocity channel maps (Fig.\,\ref{cmOIII}), the NK can be identified with the redshifted knot seen in the same maps.

Early observations by \citet{vanbreugel86} have already shown a misalignment between the radio jet south of the nucleus and the gas extension in H$\alpha$ + [N\,{\sc ii}] emission maps of $\sim 20^\circ$ (see their Fig. 7). Our observations seem to be in agreement with this evidence. The authors also observe a ``z-shaped'' structure in the gas emission maps, mainly in [O\,{\sc iii}], but on larger scales than probed in our work ($\sim 10''\,$north from the nucleus), and may not be related with the ``z-shaped'' structure we observe within the inner $\sim 3''$\, radius from the nucleus. Their lower angular resolution ($1''$) and the fact that the end part of the ``z-shaped'' structure we observe presents low surface brightness could explain why their work could not detect this structure.

\subsubsection{Line-ratio maps}
\label{sec:line_ratios}

\begin{figure*}
\centering
\includegraphics[width=\textwidth]{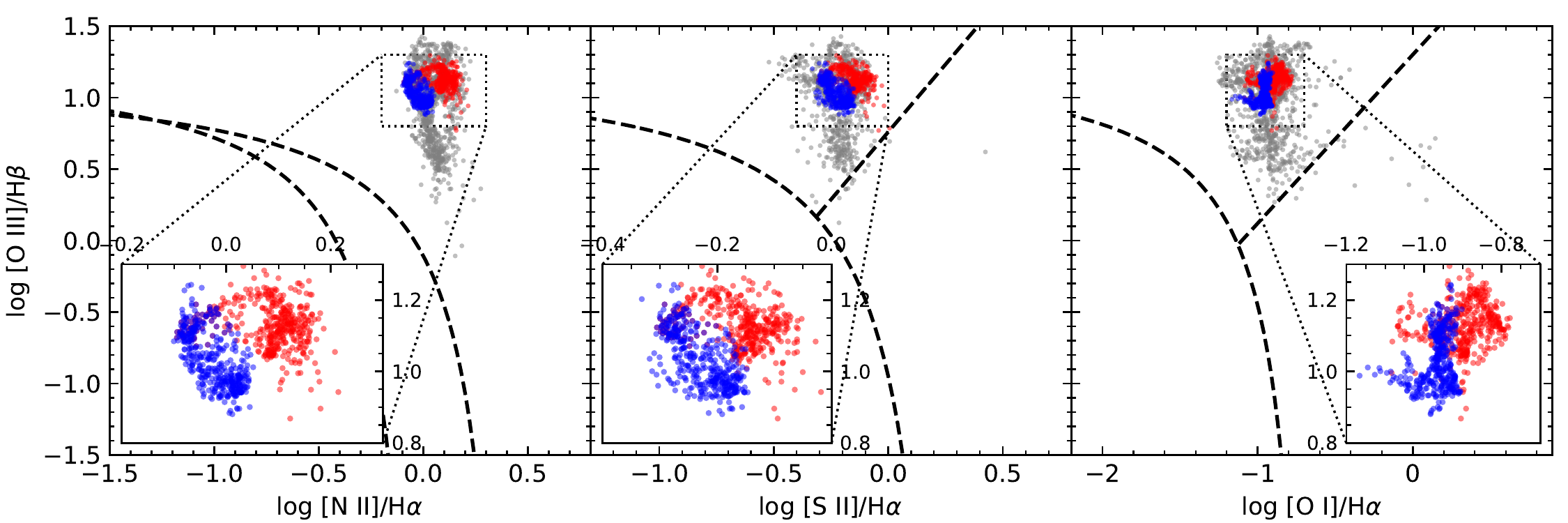}
\caption{Spatially resolved BPT diagrams of 4C\,+29.30. Each point corresponds to a single spaxel in the Gemini-GMOS datacube. Blue and red points represent spaxels located in the SK and redshifted region, respectively (see Sec.\ref{sec:out} for details on how the regions were delimited). Gray points represent other regions of the datacube. A zoomed region (dotted box) is shown for each BPT diagram to better illustrate the difference of values between the two outflowing regions. Dashed lines represent the commonly used theoretical \citep{kewley01} and empirical \citep{kauffmann03} lines that aim to separate pure star formation and star formation dominated ionization, respectively, from other ionization sources such as active nuclei.}
\label{fig:bpt}
\end{figure*}

The high values of [O\,{\sc iii}]/H$\beta$ ($0.90 < \mathrm{log ([O III]/H\beta)} < 1.11$), along with also high values of [N\,{\sc ii}]/H$\alpha$ and [S\,{\sc ii}]/H$\alpha$ ($-0.05 < \mathrm{log ([N II]/H\alpha)} < 0.11$ and $-0.22 < \mathrm{log ([S II]/H\alpha)} < -0.15$), as seen in Fig. \ref{ratios}, places the whole FoV of 4C\,+29.30 in the Seyfert region of the BPT diagrams \citep{baldwin81,kewley06}. The highest [N\,{\sc ii}]/H$\alpha$ and [S\,{\sc ii}]/H$\alpha$ ratios are observed north of the nucleus with a clear spatial correlation with the redshifted NK at 1\farcs8 from the nucleus, suggesting that this is a feature related with the AGN, possibly an outflow.  In the blueshifted region these line ratios are lower, and do not show any clear spatial correlation with the SK, which is the most characteristic feature there. Lower ratios are also observed in the north-eastern and southwestern borders of the FoV.

The [O\,{\sc i}]/H$\alpha$ ratio correlates well with the [N\,{\sc ii}]/H$\alpha$ and [S\,{\sc ii}]/H$\alpha$ maps, with the highest values associated with redshifted velocities. But in contrast to the high [N\,{\sc ii}]/H$\alpha$ and [S\,{\sc ii}]/H$\alpha$ values, they are low -- $\mathrm{log ([O I]/H\alpha)} \sim -1.0$ -- suggesting that shocks may not be a strong ionization mechanism in 4C\,+29.30. If shocks are present, they seem to be correlated with redshifted velocities.

High [O\,{\sc iii}]/H$\beta$ values, which usually trace AGN photoionization, are mostly observed north of the nucleus, $\sim 0\farcs5$ north of the dust lane. The highest values are observed $1''$ to $2''$ north-east from the nucleus, while the lowest values are observed to the south and south-west, where the gas density is somewhat higher. This suggests that the ionizing radiation of the AGN is more easily reaching the northern part of the FoV and probably being obscured in its path to ionize the gas in the southern part of the FoV. These regions of highest excitation are approximately aligned with a region where we should expect a radio counter-jet (see radio contours in Fig. \ref{large}). Although Doppler boosting is certainly an effect to consider regarding the absence of a radio counter-jet, and thus the possibility of jet-cloud interaction be the origin of this emission, the properties of the highest excitation region may indicate that it is less dense due to the fact that the gas has been ``cleared out'' by the jet, and is now prone to be photoionized by the AGN radiation, in contrast to the southern region which presents a higher density.

The BPT diagram (Fig. \ref{fig:bpt}) shows that indeed the line ratios observed in the entire Gemini-GMOS FoV lie in the Seyfert region. The difference in excitation between the SK and the redshifted region is also observed, with a clear separation between the two regions. The regions were delimited as we will further explain in Sec. \ref{sec:out}. The higher excitation shown in the spaxels located at the redshifted region in comparison to the SK indicate that indeed the dust lane is blocking much of the radiation originated from the nucleus in the southern region of the galaxy.

\subsection{Kinematics}
\label{disc_kin}

Fig. \ref{flux_velsig} and Fig. \ref{h_3_h_4} reveal that the inner few kpc of 4C\,+29.30 show multiple kinematic components in a complex scenario. Although gas rotation cannot be discarded as redshifts are observed to the north-west and blueshifts to the south-east of the nucleus, this kinematics can also be due to outflows. 

The case for outflows is supported by a number of signatures. Regarding the redshifted part of the peak velocity field: (1) as seen in Fig.\,\ref{flux_velsig}, this redshifted region is clearly discontinuous in terms of velocity to the north-east border of the FoV ($\sim 2\farcs5$ from the nucleus, to the top in Fig. \ref{flux_velsig}), where the velocity becomes zero to negative; (2) this region also presents higher velocity dispersion than the north-east border of the FoV, and  is delimited by high velocity dispersion values of $\sigma \ge 150\,$\kms; (3) the $h_4$ moment indicates the presence of more than one velocity component, as the measured negative values mean a broader peak than a Gaussian in the emission-line profile; (4) the highest redshifts are observed at the location of the NK seen, in particular, in the channel maps of Fig.\,\ref{cmOIII}; and (5) the presence of two components is also hinted from the $h_3$ map that shows a transition from negative to positive values (blue to red wings) from the south-west to north-east parts of this region.

Regarding the blueshifted part of the peak velocity field, signatures of outflows comprise: (1) the highest velocities, reaching $-200\,$\kms\, in the peak velocity map, are observed in the SK; (2) the highest velocity dispersion is also observed in this knot; (3) in the channel maps, blueshifts of up to $-600\,$\kms\, are observed in the SK region. The fact that the highest blueshifts concentrate in this knot, and the negative velocities do not increase with radius, is also a clear sign that this component is not due to rotation. Finally, the transition between the blueshifts and redshifts, in a strip $1''$ ($1.3\,$kpc) wide crossing the nucleus, approximately follows the dust lane observed in the {\it HST}-STIS image (Fig. \ref{large}) and is more abrupt than typically observed for gas rotating in a galaxy potential.

Another apparent distinct kinematic component is observed close to the north-eastern border of the FoV, with velocities close to zero. This change in the velocity field -- that is also characterized by lower velocity dispersion that drops from $\sim$ 150\,\kms\, to $\sim 80\,$\kms\, -- may be associated to the above-mentioned dust lane that bends up towards the northeast and then to the north, towards the region where zero velocities are observed. Our limited FoV does not allow us to investigate further the nature of this component.

As an additional check that the velocity field is not dominated by rotation, we have tried to fit a rotating disc model to the H$\alpha$ velocity field \citep{vanderkruit78,bertola91}, but the fit did not converge, confirming that the dominant kinematics is not ordered rotation. 

In the following sections, we present our interpretation for the structures described above.

\subsection{The southern blueshifted knot}
\label{s_knot}

The high blueshifts in the SK, of up to $\approx -600\,$\kms\, in the channel maps of Fig.\,\ref{cmOIII}, combined with the high velocity dispersion discussed above, reaching $\sim 250\,$\kms\, (Fig.\,\ref{flux_velsig}), are indications of an outflow, since the escape velocity of 4C\,+29.30 should not be higher than a few hundred \kms\, \citep{vanbreugel86}. One possibility for the origin of the outflow is an interaction between the radio jet and ambient gas, but the SK does not align well with the radio jet: while the radio jet runs along PA$\approx$200$^\circ$ (vertically in Fig.\,\ref{cmOIII}), the SK is observed 1$^{\prime\prime}$ south of the nucleus along PA$\approx$180$^\circ$ (see Fig.\,\ref{cmOIII}). 

The radio jet instead seems to be correlated with the denser gas in our observations (Fig.\,\ref{ratios}) that extends from the nucleus down to $1''$ along the jet, suggesting interaction of the jet with the surrounding medium, pushing it and increasing its density. This coincides also with part of the region covered by the dust lane. Then the jet continues outwards, possibly following a path of lowest resistance. The region where a radio counter-jet should be observed (north-east from the nucleus), in contrast, seems to display a more rarefied gas ($n_e \sim 50\,$\cmden, although we could not trace the gas density in most of this region), suggesting that most of the gas have already been ``cleared out'' by the jet, as discussed in Sec. \ref{sec:line_ratios}.

As pointed out above, the direction of the outflow, as traced by the blueshifted knot, that also corresponds to the brightest region after the nucleus, is oriented at an angle of $\approx$\,20$^\circ$ relative to the radio jet. One possibility is that the radio jet, as it progresses outwards, pushes a neighboring gas cloud partially sideways, similarly to the case of 3C\,33 \citep{couto17}, giving origin to the blueshifted knot. 

Although showing strong emission, the blueshifted knot presents emission-line ratios indicating lower excitation, what could be understood if the AGN radiation towards this knot is at least partially extincted by the dust lane. In this scenario, the ionizing radiation from the AGN would be partially hidden behind the dust lane as seen by the outflowing gas in the knot, leading to the observed low excitation.

An alternative scenario is that the SK is related to the younger milli-arcsecond radio jet observed by \citet{liuzzo09} using VLBI observations. However, the orientation of the smaller-scale radio jet differs from that of the SK by $\approx14^\circ$, and we consider this as a very unlikely scenario.

\subsection{The redshifted northern region}
\label{r_reg}

\begin{figure*}
\centering
\includegraphics[width=\textwidth]{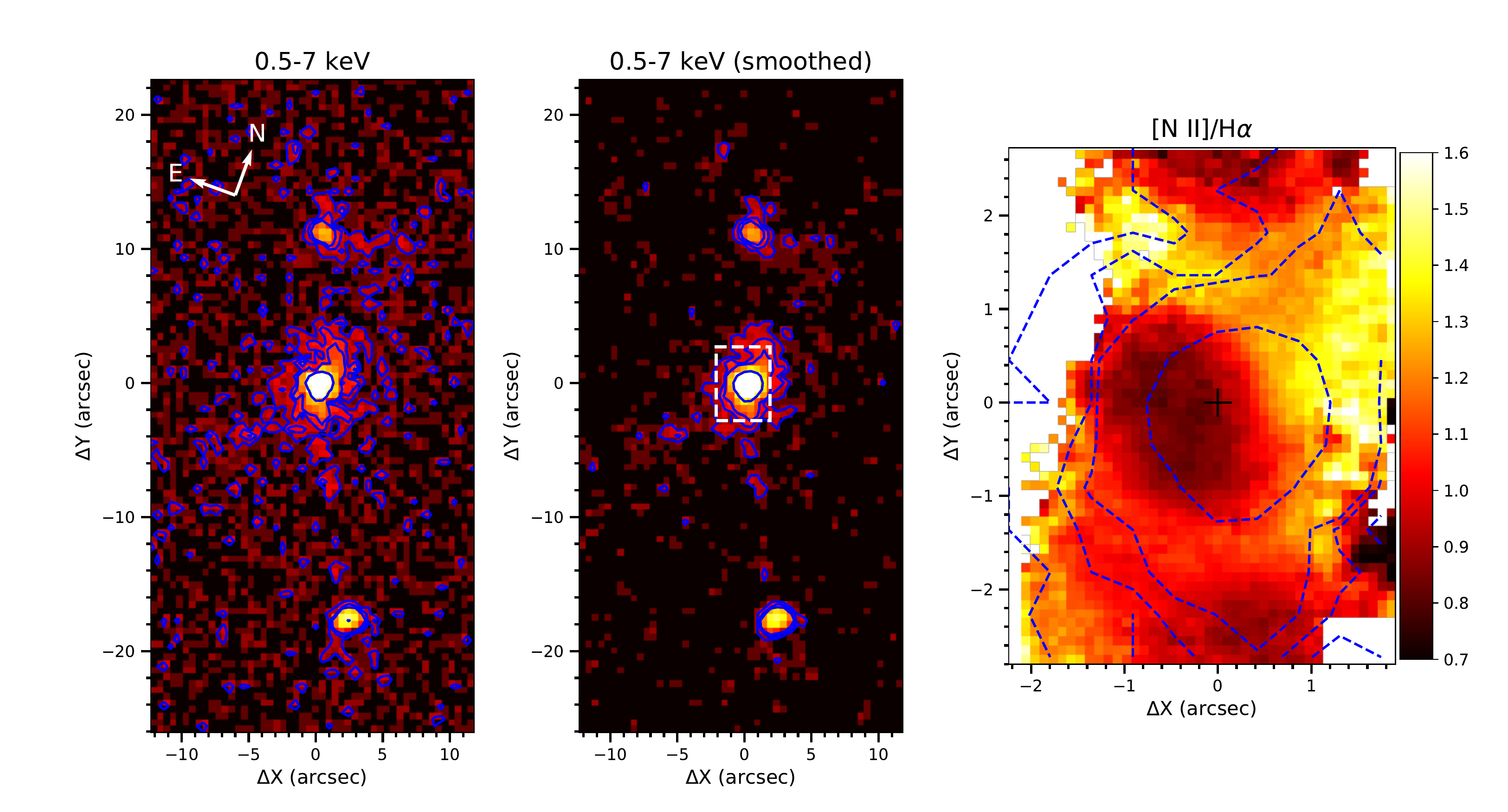}
\caption{{\it Chandra} ACIS 0.5-7 keV image of 4C\,+29.30 \citep[left panel,][]{siemiginowska12}, with a smoothed version shown in the middle panel. Smoothed X-ray emission within our Gemini-GMOS FoV (white dashed rectangle) is shown overploted the [N\,{\sc ii}]/H$\alpha$ ratio map in the right panel. X-ray contour levels are 3, 6, 10 and 100 counts.}
\label{xray}
\end{figure*}

The redshifted region centered at $\sim 1\farcs4$ to the north and north-west of the nucleus in the peak velocity maps of Fig. \ref{flux_velsig} is not as compact as the blueshifted knot and could thus be due to gas rotating in the galaxy potential. However, the decrease in the velocity values towards the north-east border of the FoV (top border in the figures), the somewhat high velocity dispersions ($\sim 130\,$\kms), the spatial correlation with the observed highest values of [N\,{\sc ii}]/H$\alpha$, [S\,{\sc ii}]/H$\alpha$ and [O\,{\sc i}]/H$\alpha$ and the suggestion of more than one kinematic component by the $h_3$ and $h_4$ moment values are all evidences that the kinematics in this region is not dominated by rotation.

The channel maps in Fig. \ref{cmOIII} show that the the highest redshifts are observed in this redshifted region in a structure that we have called the NK, at $\sim 1\farcs8$ from the nucleus and seen in the velocity channels $\sim 130-250\,$\kms. One possibility is that this redshifted knot is the counterpart of the blueshifted knot in a bipolar outflow originating in the nucleus. These two knots do indeed appear to be aligned, being at opposite sides of the nucleus, although the NK is $\approx$ 0\farcs5 farther from the nucleus than the SK. 

Although the orientation of the outflows in the redshifted and blueshifted knots deviate $\approx$\,20$^\circ$ from that of the radio jet, they share the near and far sides of the jet: the radio jet seems to be receding from us to the north-east of the nucleus, since it does not appear in the radio images, indicating Doppler boosting that enhances the south-west part. This approximate orientation is similar to that of a bipolar outflow originating the redshifted and blueshifted knots. The observed increase of line ratios in the redshifted region would be a consequence of the fact that this region receives the hard radiation from the nucleus without extinction by the dust lane, that would be in the way only of the blueshifted knot. 

The case for outflows in the blueshifted and redshifted knots is also supported by the abrupt change in velocities from blueshifts to redshifts across the strip with $\sigma \sim 170\,$\kms\, shown in Fig. \ref{flux_velsig}. Across this strip (width of $\sim 1''$), peak velocities change from $\sim -50\,$\kms\, to $\sim 100\,$\kms. This abrupt change coincides with the location of the dust lane crossing the nuclear region of the galaxy.

The left panel of Fig. \ref{xray} displays a {\it Chandra} ACIS X-ray image of 4C\,+29.30 in the 0.5-7 keV energy band. The X-ray image shows prominent emission around the nucleus and two bright spots at opposite sides of the nucleus towards the north-east and south-west that are located beyond the extent of our FoV (dashed rectangle in Fig.\,\ref{xray}). These X-ray knots are spatially correlated with radio hotspots (not shown here). The central panel shows the same image, but smoothed with a Gaussian kernel of $\sigma = 0\farcs05$. The smoothed contours help us identify extended X-ray emission towards the north by $\sim 2''$ from the nucleus then bending towards the north-east reaching out to $\sim 5''$. Our Gemini-GMOS FoV is delimited by the dashed white rectangle, and the X-ray emission in this region is shown as the blue dashed contours  over-plotted on the [N\,{\sc ii}]/H$\alpha$ ratio map in the right panel. The X-ray extended emission seems to be spatially correlated with the redshifted region that also presents high emission-line ratio values, indicating the presence of high-excitation hot gas in this region that would also be consistent with an outflow. 

Although the scenario of a bipolar outflow as the origin of  the blueshifted and redshifted knots seems to be favored by our observations, there are asymmetries in this outflow, as follows. While the apparent trajectory of the blueshifted knot makes and angle of $\sim -20^\circ$ with the radio jet axis and the knot is located at $\approx 1\farcs0$ from the nucleus, the apparent trajectory of the redshifted knot deviates by $\sim -35^\circ$ from the radio jet axis and the knot is located $\approx 1\farcs4$ from the nucleus. The highest velocities relative to the nucleus are also not symmetric, with the SK presenting absolute velocities at least $\approx 50\,$\kms\, higher than that of the NK. On the other hand, one can argue that these asymmetries could be due to density asymmetries in the surrounding gas, and we thus consider the scenario of a bipolar outflow a plausible one.

\subsection{The dust lane}

\begin{figure}
\centering
\includegraphics[width=\columnwidth]{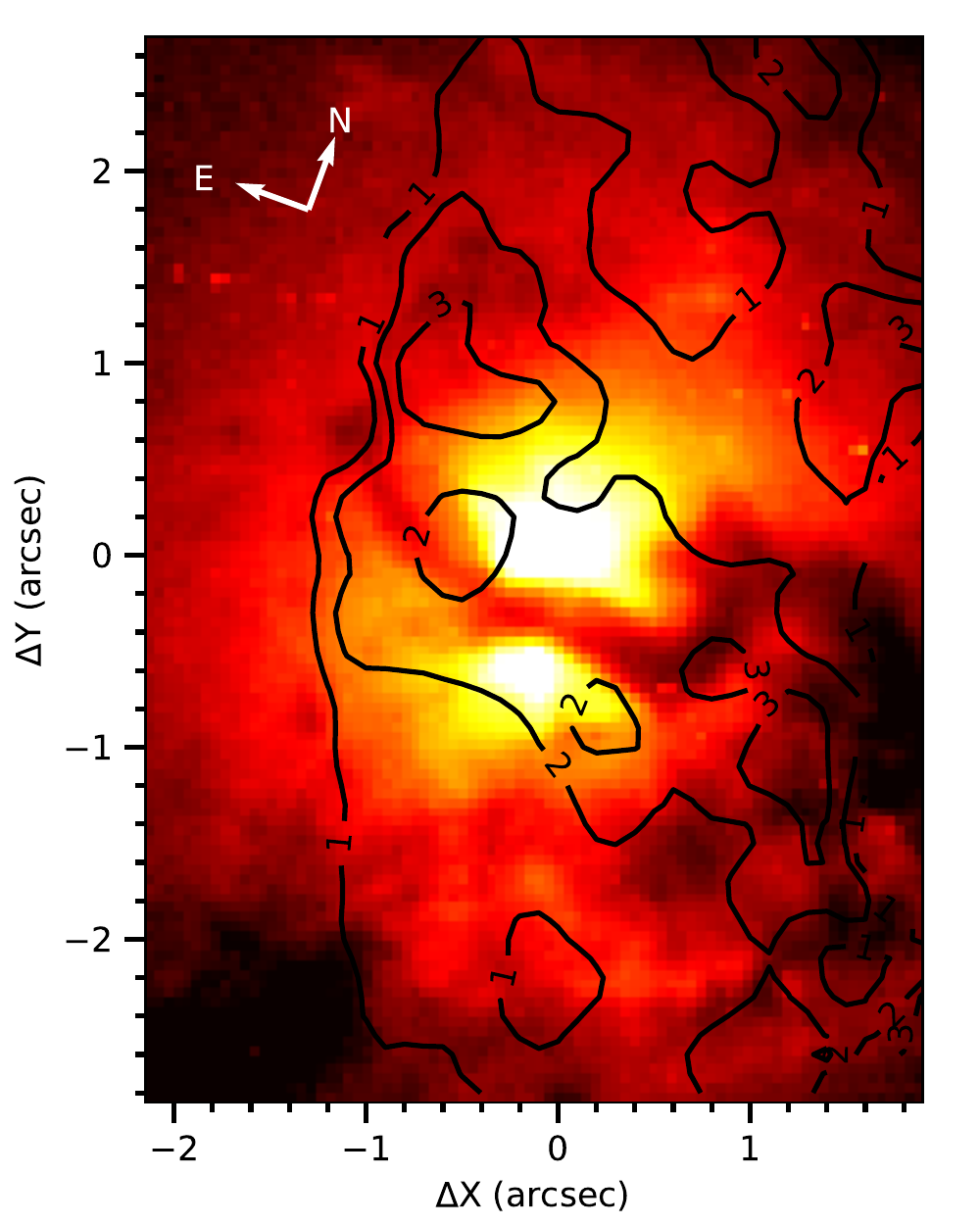}
\caption{{\it HST}-STIS optical image of 4C\,+29.30, as shown in Fig. \ref{large}, but within GMOS-IFU FoV, shown also with $A_V$ contours from the bottom central panel of Fig. \ref{ratios}.}
\label{red}
\end{figure}

The dust lane clearly affects a large region of the FoV we are probing in this study and must be considered in our discussions. Fig. \ref{red} shows the {\it HST}-STIS image of 4C\,+29.30 within the FoV of our GMOS data with the contours of the $A_V$ optical extinction map from Fig. \ref{ratios} superimposed. As expected, there is a spatial correlation between the dust lane and the highest values of $A_V$. Extinction of $> 2\,$mag is observed in the region covered by the dust lane, reaching $\sim 4\,$mag at a few locations.

The dust lane seems to affect more the southern part of our FoV. The {\it HST}-STIS image shows higher emission to the north when compared to the south. The [N\,{\sc ii}]/H$\alpha$, [S\,{\sc ii}]/H$\alpha$, [O\,{\sc i}]/H$\alpha$ and [O\,{\sc iii}]/H$\beta$ line ratios, which trace the impact of the AGN radiation in the ISM, all present higher values in the north than in the south. The [O\,{\sc iii}]/H$\beta$ ratio shows an increase just above the border of the dust lane. Also, the gas density is higher to the south of the nucleus, but also partially coincident with the dust lane. We thus interpret that most of these characteristics are linked to the dust lane.

The dust lane crosses the central region of 4C+29.30, resembling the case of Centaurus A, although the dust lane appears to be more distorted than that in Centaurus A. We know that it is the signature of a past merger event in Centaurus A, as it seems also to be the case of 4C\,+29.30 \citep{siemiginowska12}.

\subsection{Additional kinematic components}

Another couple of kinematic components, which may be related to each other, are two emitting regions observed at the top and bottom borders of our FoV, at $\gtrsim 2\farcs0$ from the nucleus, to the north-east (top of the FoV) and south-west (bottom). These regions are most clearly observed in the channel maps with velocities close to zero (Fig. \ref{cmOIII}), making the flux distribution maps to present a ``z-shaped'' structure between the channel maps $-85\,\ge v \ge 135\,$\kms. Although these regions may seem connected to the more internal SK and NK, the knots show their strongest contribution at higher velocity channels. This can also be observed in the kinematic maps of Fig.\ref{flux_velsig} where we observe a decrease in the peak velocity towards the borders of the FoV. The velocity dispersion also decreases towards the borders of the FoV and in particular to $\sigma \sim 70\,$\kms\, at the north-eastern (top) border, supporting the existence of only one kinematic component there. Although showing low velocities, the fact that the bottom part of the `Z' is mostly observed in blueshift channels and the upper part in redshift channels suggest that this kinematics could be due to overall mild rotation of the ambient gas. 

Still another component can be observed in the channel maps of Fig. \ref{cmOIII} for velocities $-197\,\ge v \ge 25\,$\kms: a region of increased gas emission at the south-western border of the FoV that correlates with a knot seen in the radio jet, possibly the result of jet-gas interaction. Another region at the north-eastern border of the FoV in the channel map with $v=25\,$\kms, is also partially seen, aligned in the direction of the radio jet. This could be a counterpart structure to the south-western region resulting from the interaction of the ambient gas with the north-eastern counterpart of the radio jet. Nevertheless, the alignment and spatial correlation with the radio knot for the south-western region seems to be the only evidence of interaction, as emission in these regions is only observed at low absolute velocities ($v < 200\,$\kms) and the velocity dispersion does not increase. Actually it decreases at the northern region, with $\sigma < 100\,$\kms. The line ratios also decrease in these regions, as observed in Fig. \ref{ratios}.

We can compare the scenario we propose in our work with the results obtained by \citet{vanbreugel86}, where the gas kinematics and excitation of 4C\,+29.30 were also studied. Our data set is quite different from this work: while we obtain IFS data of the inner $\sim 3\farcs5 \times 5''$ with a $\approx 0\farcs77$ spatial resolution, \citet{vanbreugel86} present optical imaging of a FoV up to $130'' \times 130''$ with spatial resolutions of $1''$ (red broad band and H$\alpha$) and $1\farcs5$ (H$\alpha$ and [O\,{\sc iii}]) and long slit spectroscopy with $1''$ angular resolution (along P.A. $= 4^\circ$ and $95^\circ$).

Similarly to our results, \citet{vanbreugel86} analysis based on the emission-line gas adjacent to the radio emission regions indicated the presence of jet-gas interactions. However, their data covered larger scale structures, up to $\sim 20''$ from the nucleus, while we are able to probe nuclear regions within the central $< 3 \arcsec$ radius. Additionally, the $\sim 20^\circ$ misalignment between the radio and the optical emission reported in their analysis is similar to the one we observed between the radio jet and the SK (see Sec. \ref{s_knot}). Thus we are able to observe in detail the beginning of the jet-gas interactions that extends much farther than our FoV.

In summary, we conclude that the gas kinematics indicates the presence of a bipolar outflow observed as the SK and NK, plus an underlying component showing a hint of rotation, what can only be confirmed with observations over a larger FoV. There is also evidence of jet-cloud interactions at the top and bottom borders of the FoV in regions located north-east and south-west from the nucleus.



\subsection{Estimates of the ionized gas physical properties}

\subsubsection{Ionized gas mass}
\label{sec:gas_mass}

\begin{figure}
\centering
\includegraphics[width=\columnwidth]{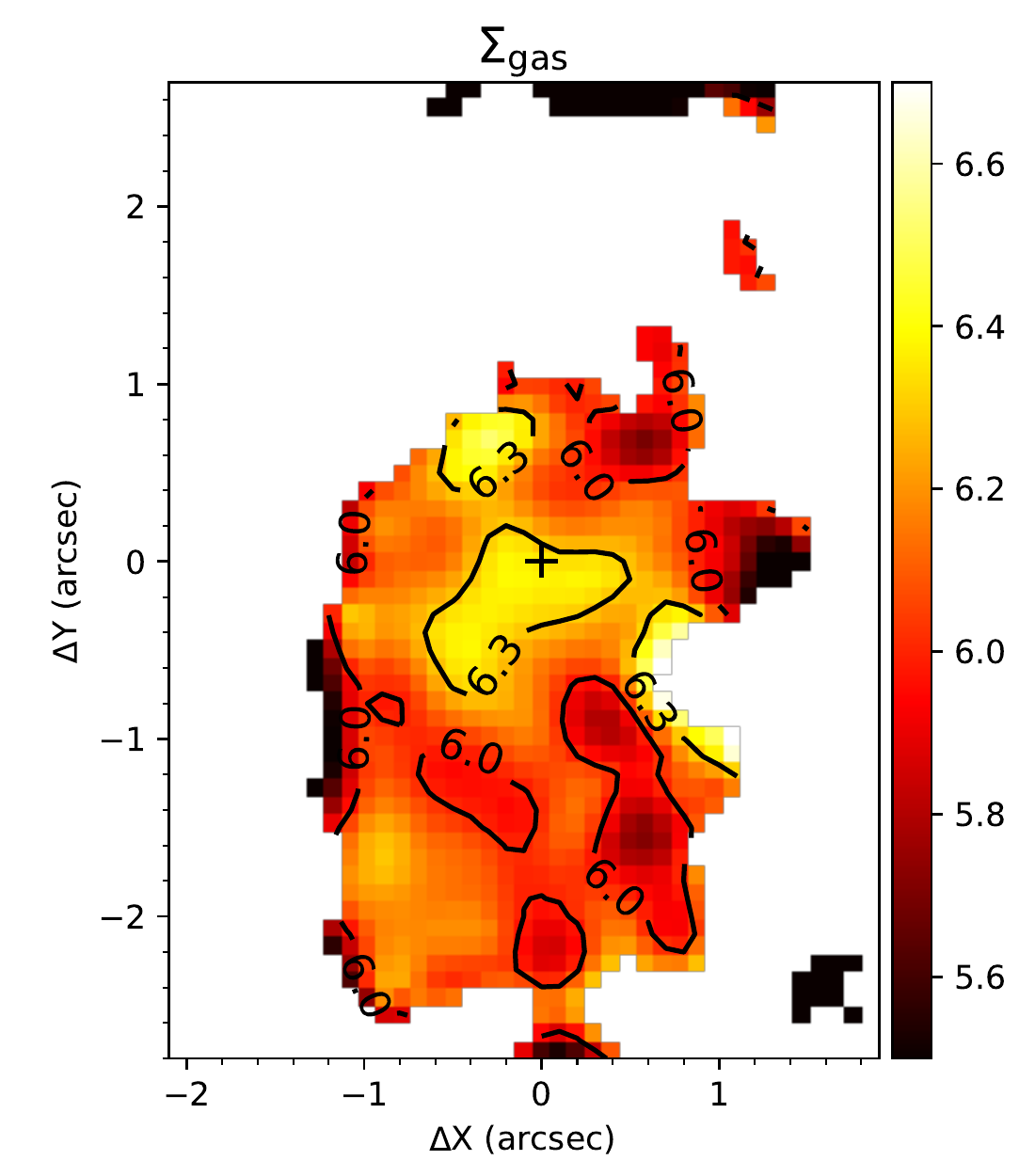}
\caption{Ionized gas mass distribution in the inner few kpc of 4C\,+29.30. Units are $M_\odot\, \mathrm{arcsec}^{-2}$ and are in logarithmic scale.}
\label{gas_mass}
\end{figure}

We can estimate the total mass of the emitting ionized hydrogen gas as:

\begin{equation}
M_{\mathrm{gas}} \approx 2.3 \times 10^5\,{\frac{L_{41}(\textrm{H}\alpha)}{n_3}}\,M_\odot \, , 
\end{equation}

\noindent
where $L_{41}(\textrm{H}\alpha)$ is the H$\alpha$ luminosity in units of $10^{41}\,$\ergs\, and $n_3$ is the electron density in units of $10^3$\,cm$^{-3}$. To calculate the H$\alpha$ luminosity we corrected the emitted flux for reddening assuming the \citet{cardelli89} reddening law, with $R_V = 3.1$. Fig. \ref{gas_mass} shows the resulting ionized gas mass distribution. With high uncertainties in the H$\beta$ flux and [S\,{\sc ii}] ratio (and consequently in the electron density), we had to mask part of the map, leaving it restricted to mainly the nucleus and the SK, the regions where the emission is strongest. In any case, due to the lower luminosity, the contribution from the other regions should be much lower. No particular structure is observed in the gas mass map, except for showing a modestly larger mass to the north-east (top) of the FoV as compared to the south-west (bottom), as expected from the strongest emission of the former.

We obtain a total H$\alpha$ luminosity of $L(\textrm{H}\alpha) = 1.4 \pm 0.8 \times 10^{42}\,$\ergs and a total ionized hydrogen gas mass of $M_{\mathrm{gas}} = 1.6 \pm 0.8 \times 10^7\,$M$_\odot$ within the $\sim 3\,$kpc radius of 4C\,+29.30 probed by our data. Mass uncertainties are derived only considering the uncertainties from the flux measurements, and then reproduced in the luminosity derivation. This value is in agreement with the $4'' \times 4''$\,nuclear aperture ionized gas mass obtained by \citet{vanbreugel86}, of $M_{\mathrm{gas}} = 1.2 \times 10^7\,$M$_\odot$ (see their Table 7B).

\subsubsection{Mass outflow rate and outflow kinetic power}
\label{sec:out}

As discussed in Sec. \ref{s_knot}, the SK, located $\sim 1\farcs0$ south of the nucleus, presents high blueshifts ($\approx -600\,$\kms) associated with an increase in the velocity dispersion ($\sim 250\,$\kms). The redshifted NK at $\sim 1\farcs4$ from the nucleus, described in Sec. \ref{r_reg}, could be a counterpart to the SK. It is characterized by high excitation and seems to correlate with warm gas emitting X-rays. This region also presents high redshifts ($\approx 550\,$\kms), as seen in the channel maps of Fig. \ref{cmOIII}.

We estimate the mass outflow rate in the SK and NK by adopting a  biconical geometry for the outflowing gas, as one side is approaching and the other is receding relative to the line of sight. Also, we adopt a non-zero inclination between the plane of the sky and the bicone axis, since we observe high redshifts and blueshifts in regions that we identify with the end (base) of the cones. We consider that both cones have a base with diameter of $1\farcs0$ (in agreement with the observed geometry of the emission in Fig. \ref{cmOIII}), but with different projected heights, $1\farcs0$ for the SK and $1\farcs4$ for the NK.

The mass outflow rate can be calculated as:

\begin{equation}
\dot{M}_{out} = 1.4\,n_e\,m_p\,v_{out}\,A\,f\, ,
\label{mout}
\end{equation}

\noindent
where $m_p = 1.7 \times 10^{-24}\,$g is the proton mass, $n_e$ is the electron density, $v_{out}$ is the velocity of the outflow perpendicular to $A = \pi\,r^2$, which is the cross section (base) of the cone, $f$ is the filling factor, and the factor 1.4 is to account for elements heavier than hydrogen. Assuming the geometry mentioned above, we obtain a cross section area of the outflow of $A = 4.6 \times 10^{43}\,$cm$^2$ for both outflowing regions (SK and NK). The filling factor can be obtained under the assumption that the H$\alpha$ emission corresponds to case B recombination \citep{osterbrock06}:

\begin{equation}
f = 2.6 \times 10^{59}\, {\frac{L_{41}(\textrm{H}\alpha)}{V\,n_3^2}}\, ,
\label{eq_f}
\end{equation}

\noindent
where $V$ is the volume of the emitting region in $\rm cm^{3}$ and $n_3$ is the electron density in units of $10^3$\,$\rm cm^{-3}$. These calculations and assumptions are described with more details in \citet{peterson97}.

To delimit the outflowing regions in the FoV, we have used 
flux contours from the \o3 channel maps in Fig. \ref{cmOIII} in order to create two masks. The adopted flux contour corresponds to the value of $1 \times 10^{-17}\,$\ergcmspa, and we have used the channel map with velocity $-308\,$\kms\, to delimit the region of blueshifted outflows in the SK, and $246\,$\kms\, to delimit the region of redshifted outflows in the NK. We then considered outflowing the gas in the channels with velocities more negative than $-308\,$\kms\, for the blueshifted outflows and with velocities more positive than $246\,$\kms\, for the redshifted outflows. 

The channel maps in Fig. \ref{cmOIII} are for the \o3 line. We have decided to use \o3 instead of the H$\alpha$ channel maps because the H$\alpha$ emission line is blended with the \n2 line at high velocities, which are exactly the ones we need to use in our calculations. To convert \o3 to H$\alpha$ emission, we used the measured \o3 /H$\alpha$ ratio from our emission-line fits for each spaxel, and then correcting for dust extinction. Although this ratio does not necessarily correspond to the ratio in each channel map, this ratio does not vary much, so that the small variations do not have much impact on the resulting integrated flux. As the [SII] line ratio does not seem to vary much along the line profiles, and we would not have a good signal-to-noise ratio to try to build a ``density channel map", we have decided to use a mean density value for each knot region. Using these values, we calculated the filling factors for each channel and for each outflowing knot.

Finally, we calculated the mass outflow rate for each channel and knot region, using eq. \ref{mout}, deprojecting the velocities according to the adopted inclinations of the biconical outflow. We then calculated the kinetic power of the outflow as \citep{holt06,mahony16}:

\begin{equation}
\dot{E} = 6.34 \times 10^{35} \, {\frac {\dot{M}_{out}}{2}} \, (v_{out}^2 + 3\sigma^2) .
\label{eq_Ekin}
\end{equation}

\noindent
where we used the mean velocity dispersion for each knot region.

We show in Table \ref{tab:cms} the values of the calculated properties above for each knot and each channel considered to correspond to the outflows and for three different inclinations: H$\alpha$ luminosities, ionized gas masses (as estimated in Sec. \ref{sec:gas_mass}), deprojected channel map velocities, filling factors, mass-outflow rates and outflow kinetic powers. Notice that the H$\alpha$ luminosities and gas masses do not change with the inclination assumed, since they only depend on the integrated line fluxes in the knot regions.

\begin{table*}
\caption{Physical properties of the outflowing ionized gas in 4C\,+29.30, separated by channel maps velocities and cone inclinations.}
	\label{tab:cms}
	
\begin{tabular}{|l|c|c|c|c|c|c|c|c|}
\hline
\hline

\begin{tabular}{c}
\o3 channel maps velocity (\kms)
\end{tabular}&
$-641$ & $-530$ & $-419$ & $-308$ & $246$ & $357$ & $468$ & $579$ \\

\hline

$L(\textrm{H}\alpha)$ ($\times 10^{41}$ \ergs)&
$0.48$ & $0.91$ & $1.43$ & $2.13$ & $1.52$ & $0.64$ & $0.34$ & $0.27$ \\ 

$M_{\mathrm{gas}}$ ($\times 10^{6}$ M$_\odot$)&
0.6 & 1.1 & 1.8 & 2.7 & 3.1 & 1.3 & 0.7 & 0.5 \\ 

\hline
\multicolumn{7}{c}{Cone inclination $= 40^\circ$}\\
\hline

Deprojected $v_{\textrm{out}}$ (\kms)&
$-997$ & $-824$ & $-652$ & $-479$ & $383$ & $555$ & $728$ & $901$ \\ 

$f$ ($\times 10^{-4}$)&
0.5 & 0.9 & 1.5 & 2.2 & 3.0 & 1.2 & 0.7 & 0.5 \\ 

$\dot{M}_{\mathrm{out}}$ (\msunyr)&
1.5 & 2.4 & 3.0 & 3.3 & 2.2 & 1.3 & 0.9 & 0.9 \\ 

$\dot{E}$ ($\times 10^{41}$ \ergs)&
5.3 & 5.8 & 4.8 & 3.3 & 1.5 & 1.6 & 1.8 & 2.5 \\ 

\hline
\multicolumn{7}{c}{Cone inclination $= 30^\circ$}\\
\hline

Deprojected $v_{\textrm{out}}$ (\kms)&
$-1282$ & $-1060$ & $-838$ & $-616$ & $492$ & $714$ & $936$ & $1158$ \\ 

$f$ ($\times 10^{-4}$)&
0.6 & 1.0 & 1.6 & 2.4 & 3.4 & 1.4 & 0.8 & 0.6 \\ 

$\dot{M}_{\mathrm{out}}$ (\msunyr)&
2.2 & 3.5 & 4.4 & 4.8 & 3.2 & 1.9 & 1.4 & 1.3 \\ 

$\dot{E}$ ($\times 10^{41}$ \ergs)&
12.3 & 13.5 & 10.9 & 7.0 & 3.2 & 3.5 & 4.1 & 5.9 \\ 

\hline
\multicolumn{7}{c}{Cone inclination $= 50^\circ$}\\
\hline

Deprojected $v_{\textrm{out}}$ (\kms)&
$-837$ & $-692$ & $-547$ & $-402$ & $321$ & $466$ & $610$ & $756$ \\ 

$f$ ($\times 10^{-4}$)&
0.4 & 0.8 & 1.2 & 1.8 & 2.5 & 1.0 & 0.6 & 0.4 \\ 

$\dot{M}_{\mathrm{out}}$ (\msunyr)&
1.1 & 1.7 & 2.1 & 2.3 & 1.5 & 0.9 & 0.7 & 0.6 \\ 

$\dot{E}$ ($\times 10^{41}$ \ergs)&
2.7 & 3.0 & 2.6 & 1.8 & 0.8 & 0.8 & 0.9 & 1.3 \\
\hline

\end{tabular}
\end{table*}

\begin{table}
\caption{Integrated physical properties of the outflowing ionized gas in 4C\,+29.30, separated by the outflowing regions and cone inclinations.}
	\label{tab:tots}
	
\setlength{\tabcolsep}{3.5pt}
\begin{tabular}{|l|c|c|c|}
\hline
\hline

& 
\begin{tabular}{c}
SK
\end{tabular}&
\begin{tabular}{c}
NK
\end{tabular}&
Total \\

\hline

$L(\textrm{H}\alpha)$ ($\times 10^{41}$ \ergs)&
$4.9$ & $2.8$ & $7.7$  \\ 

Mean $n_e$ ($\rm cm^{-3}$)&
181.9 & 111.0 & 136.8  \\ 

$M_{\mathrm{gas}}$ ($\times 10^{6}$ M$_\odot$)&
6.2 & 5.7 & 12.0  \\ 

Mean $\sigma$ (\kms)&
166.5 & 152.2 & 156.9  \\ 

\hline
\multicolumn{4}{c}{Cone inclination $= 40^\circ$}\\
\hline

$V$ ($\times 10^{65} \rm cm^{3}$)&
0.77 & 1.1 & 1.8  \\  

$f$ ($\times 10^{-4}$)&
5.1 & 5.4 & 5.8  \\  

$\dot{M}_{\mathrm{out}}$ (\msunyr)&
15.9 & 9.4 & 25.4  \\  

$\dot{E}$ ($\times 10^{42}$ \ergs)&
5.4 & 2.6 & 8.1  \\  

\hline
\multicolumn{4}{c}{Cone inclination $= 30^\circ$}\\
\hline

$V$ ($\times 10^{65} \rm cm^{3}$)&
0.68 & 0.95 & 1.6  \\  

$f$ ($\times 10^{-4}$)&
5.7 & 6.1 & 6.6  \\  

$\dot{M}_{\mathrm{out}}$ (\msunyr)&
23.2 & 13.7 & 36.9  \\  

$\dot{E}$ ($\times 10^{42}$ \ergs)&
12.7 & 6.1 & 18.8  \\  

\hline
\multicolumn{4}{c}{Cone inclination $= 50^\circ$}\\
\hline

$V$ ($\times 10^{65} \rm cm^{3}$)&
0.91 & 1.3 & 2.2  \\  

$f$ ($\times 10^{-4}$)&
4.2 & 4.6 & 4.9  \\  

$\dot{M}_{\mathrm{out}}$ (\msunyr)&
11.2 & 6.6 & 17.9  \\  

$\dot{E}$ ($\times 10^{42}$ \ergs)&
2.8 & 1.3 & 4.1  \\  

\hline

\end{tabular}
\end{table}

Initially we assume an inclination angle of $40^\circ$ between the bicone axis and the plane of the sky, considering that this inclination should not differ much from that of the radio jet, as determined by \citet{liuzzo09} in their study of the radio jet with milliarcsec resolution VLBA images. We also made estimates for inclinations $30^\circ$ and $50^\circ$, in order to evaluate how much the calculated properties vary if the uncertainty in the inclination is of the order of $10^\circ$. The results of these calculations are shown in Table \ref{tab:cms}, while integrated values for the outflow parameters, for both the SK and NK, as well as the two combined are shown in Table \ref{tab:tots}.

We obtain mass outflow rates of $\dot{M}_{out\, \textrm{SK}} = 15.9 \substack{+7.3 \\ -4.7} \,$\msunyr\, and $\dot{M}_{out\, \textrm{NK}} = 9.4 \substack{+4.3 \\ -2.8} \,$\msunyr\, for each separate knot, with a kinetic power of $\dot{E}_{out\, \textrm{SK}} = 5.4 \substack{+7.3 \\ -2.6}\, \times 10^{42}\,$\ergs\, and $\dot{E}_{out\, \textrm{NK}} = 2.6 \substack{+3.5 \\ -1.3}\, \times 10^{42}\,$\ergs, using the most likely inclination of $40^\circ$ and considering the uncertainties as solely due to the variation of $\pm 10^{\circ}$ in the inclination. This gives us a total mass outflow rate of $\dot{M}_{out} = 25.4 \substack{+11.5 \\ -7.5}\,$\msunyr, with an outflow kinetic power of $\dot{E} = 8.1 \substack{+10.7 \\ -4.0}\, \times 10^{42}\,$\ergs. In comparison, ionized gas mass outflow rates of Seyfert galaxies are usually in the range $0.1$--$10\,$\msunyr \citep{veilleux05}, indicating that in 4C\,+29.30 the outflows show higher mass loads of ionized gas than typically observed in other nearby active galaxies of similar luminosity. We also note that we obtain a total filling factor of $f = 5.8 \times 10^{-4}$, at least one order of magnitude lower than usually obtained for narrow line regions (NLRs) of Seyfert galaxies \citep[$\sim 10^{-1}$--$10^{-3}$,][]{osterbrock06,storchi10}, suggesting that the outflows we observe in 4C\,+29.30 are more ``rarefied'', although comparable to other similar studies in radio galaxies \citep[e.g. 4C\,12.50,][]{holt11}. 

It is important to note that the calculations presented above are based on a number of assumptions and the uncertainties may be larger than quoted above. The main source of uncertainty is probably the outflow geometry, which is based on the projected geometry in the plane of the sky. A small change in the geometry can have considerable influence in the calculated properties of the outflow. The quoted uncertainties correspond to different bicone inclinations; the resulting values of the relevant parameters are given in Tables \ref{tab:cms} and \ref{tab:tots}. And there are other uncertainties that can be considered, such as in the gas density, that we have adopted as derived from the [S{\sc ii}] ratios, while recent studies suggest may be higher in the outflows when estimated using transauroral lines \citep{holt11,harrison18,santoro18,baron19,davies20}.

Assuming that the bolometric luminosity of the AGN is $L_{\textrm{bol}} \sim 100 \times L(\textrm{H}\alpha)$, we have that $L_{\textrm{bol}} = 1.4 \pm 0.8 \times 10^{44}\,$\ergs from the $L(\textrm{H}\alpha)$ obtained in the previous section. This means that the total outflow kinetic power represents $5.8 \substack{+7.6 \\ -2.9} \%$ of the bolometric luminosity, considering the case for the cone inclination of $40^\circ$. This result indicates that the outflow in 4C+29.30 is powerful enough to have a significant impact in the evolution of the host galaxy, according to models \citep[e.g.][]{hopkins10}.

This agrees with the values we have obtained for $\dot{M}_{out} = 25.4\,$\msunyr, and $\dot{E} = 8.1 \times 10^{42}\,$\ergs, when compared to $L_{\textrm{bol}}=1.4 \pm 0.8 \times 10^{44}\,$\ergs, put 4C+29.30 above the values obtained by \citet{fiore17} for $\dot{M}_{out}$ vs. $L_{\textrm{bol}}$ ($\dot{M}_{out} \sim 0.1\,$\msunyr\, for $L_{\textrm{bol}} \sim 10^{44}\,$\ergs) and $\dot{E}$ vs. $L_{\textrm{bol}}$ ($\dot{E} < 10^{42}\,$\ergs\, for $L_{\textrm{bol}} \sim 10^{44}\,$\ergs) in ionized gas outflows in AGNs. The ionized gas mass outflow rate and the outflow kinetic power in 4C\,+29.30 are also higher than those estimated in other radio galaxies such as PKSB 1934-63 \citep{santoro18}, 3C\,293 \citep{mahony16}, PKS 1345+12 \citep{holt11}, 3C\,33 \citep{couto17} and Arp\,102B \citep{couto13}, but with a kinetic power comparable to that obtained for ESO428-G14 \citep{may18}. However, one should take into account that this comparison may not be true due to different range in source sizes and different estimation methodologies, and more parameters would be needed to do a proper comparison.

Also, considering the mass accretion rate of the SMBH to be $\dot{m} = L_{\textrm{bol}} / \eta\,c^2$, where $\eta$ is the accretion efficiency usually assumed to be 0.1, we obtain $\dot{m} = 0.024 \pm 0.01\,$\msunyr, $\sim754$ times lower than the total mass outflow rate, confirming that the outflow can only be due to mass loading of a nuclear outflow. We also note that this ratio value is $\sim 3$ times higher than typical values in the literature for AGNs with powerful outflows \citep[e.g.,][]{bae17}.

\section{Conclusions}
\label{conc}

We have studied the excitation and kinematic properties of the ionized gas of the inner $4.3\,\times\, 6.2\,$kpc$^2$ of the interacting radio galaxy 4C\,+29.30 using Gemini-GMOS integral field spectroscopy.

The main conclusions of this work are:
\begin{enumerate}

\item{Flux maps of the [O\,{\sc iii}] and H$\alpha$ emission lines show evidence of the interaction observed in previous studies, in a ``z-shaped'' extended emission morphology, with the highest emission, aside from the nucleus, being observed $\sim 1''$ south from the nucleus, in a region we have called the ``southern knot (SK)'';}

\item{The SK presents high blueshifts (peak velocities of $\sim 200\,$\kms) and high velocity dispersions ($\sim 250\,$kms). [O\,{\sc iii}] channel maps show emission at negative velocities up to $\sim -650\,$\kms\, in this region, strongly indicating the presence of an outflow there;}

\item{A possible redshifted counterpart to the SK is observed $\sim 1\farcs4$ north from the nucleus, in the ``northern knot (NK)''. Although absolute velocities are $\approx20\%$ lower than in the SK (both for the peak velocities and in the  channel maps), we conclude that the most probable scenario is that the SK and NK are the product of a bipolar outflow, possibly due to jet-gas interaction;}

\item{We obtain the highest electron densities ($\sim 400$\,\cmden) in a region spatially correlated with the radio jet extended from the nucleus down to $1''$ south-west of it, in the middle of the conspicuous dust lane crossing the brightest part of the galaxy, just above the SK. This correlation supports the presence of interactions between the radio jet and the surrounding medium;} 

\item{The presence of jet-cloud interaction is also observed in an emission knot seen in the [O\,{\sc iii}] channel maps close to zero velocity at $\sim 3''$ south-southwest of the nucleus that coincides with a radio hotspot;}

\item{The dust lane causes the highest extinction (reaching up to $A_V > 3.0$) to be observed in a $\sim 1\farcs5$ wide strip passing through the nucleus along the east-west direction of our FoV;}

\item{The gas excitation is higher in the redshifted region to the north, that is also correlated with X-ray emission, suggesting that gas is warmer there. We attribute this difference between the north and south to the possibility that the ionizing radiation has a clearer path to the north than to the south, due to the dust lane that seems to partially block the nuclear radiation to the blueshifted region in the south;}

\item{Although not seeming spatially correlated with the radio jet, we suggest that jet-cloud interaction could be the origin of the outflows in the SK and NK, that may be gas clouds pushed aside and outwards by the passage of the radio jet;}

\item{Considering that both the SK and NK are tracing gas outflows, we obtain estimates for the mass outflow rates of $\dot{M}_{out\, \textrm{SK}} = 15.9 \substack{+7.3 \\ -4.7} \,$\msunyr and $\dot{M}_{out\, \textrm{NK}} = 9.4 \substack{+4.3 \\ -2.8} \,$\msunyr, with a total of $\dot{M}_{out} = 25.4 \substack{+11.5 \\ -7.5} \,$\msunyr. The total outflow kinetic power represents $5.8 \substack{+7.6 \\ -2.9} \%$ of the bolometric luminosity ($L_{\textrm{bol}} = 1.4 \pm 0.8 \times 10^{44}\,$\ergs).}

\indent

The mass outflow rates and powers above are in approximate agreement with previous values in the literature, but for more luminous AGN than 4C\,+29.30. We find higher values for the mass outflow rate and power than those predicted by the relation between these quantities and $L_{\textrm{bol}}$ obtained by \citet{fiore17}. This means higher values of the ratio between mass outflow rate and accretion mass rate than the expected. The ratio between the outflow power and $L_{\textrm{bol}}$ of $\approx$\,4\% implies that the outflows in 4C\,+29.30 can cause an important impact on the evolution of the host galaxy \citep[according to][a kinetic power of $\dot{E} \sim 0.5\% L_{\textrm{bol}}$ could be enough to drive a considerable amount of gas outwards]{hopkins08}. 4C\,+29.30 presents more powerful outflows than two radio galaxies previously studied by our group \citep[Arp 102B and 3C\,33,][]{couto13,couto17}, suggesting that this could be due to the jet reactivation and young radio emission, in agreement with other studies in compact radio sources \citep{holt08,molyneux19}.

\end{enumerate}

\section*{Acknowledgments}

We would like to thank the anonymous referee for their careful comments which lead to a improved paper. GSC acknowledges the support by the Comit\'{e} Mixto ESO-Chile and the DGI at Universidad de Antofagasta (CR 4731), from CONICYT FONDECYT project No. 3190561, and from the Brazilian institutions CNPq and CAPES. RAR  acknowledges support from  Conselho Nacional de Desenvolvimento Cient\'ifico e Tecnol\'ogico (302280/2019-70) and Funda\c c\~ao de Amparo \`a pesquisa do Estado do Rio Grande do Sul (17/2551-0001144-9 and 16/2551-0000251-7). AS work was supported by NASA contract NAS08-03060 9 Chandra X-ray Center).

Based on observations obtained at the Gemini Observatory (processed using the Gemini IRAF package), which is operated by the Association of Universities for Research in Astronomy, Inc., under a cooperative agreement with the NSF on behalf of the Gemini partnership: the National Science Foundation (United States), National Research Council (Canada), CONICYT (Chile), Minist\'{e}rio da Ci\^{e}ncia, Tecnologia e Inova\c{c}\~{a}o (Brazil) and Ministerio de Ciencia, Tecnolog\'{i}a e Innovaci\'{o}n Productiva (Argentina), and Korea Astronomy and Space Science Institute (Republic of Korea). {\sc IRAF} is the Image Reduction and Analysis Facility, a general purpose software system for the reduction and analysis of astronomical data. 

{\sc IRAF} is written and supported by the National Optical Astronomy Observatories (NOAO) in Tucson, Arizona. NOAO is operated by the Association of Universities for Research in Astronomy (AURA), Inc. under cooperative agreement with the National Science Foundation. 

Based on observations made with the NASA/ESA Hubble Space Telescope, obtained from the data archive at the Space Telescope Science Institute. STScI is operated by the Association of Universities for Research in Astronomy, Inc. under NASA contract NAS 5-26555. 

The National Radio Astronomy Observatory is a facility of the National Science Foundation operated under cooperative agreement by Associated Universities, Inc. 

Based on observations made with the NASA/ESA Hubble Space Telescope, obtained from the data archive at the Space Telescope Science Institute. STScI is operated by the Association of Universities for Research in Astronomy, Inc. under NASA contract NAS 5-26555. 

This research has made use of data obtained from the Chandra Data Archive, and software provided by the Chandra X-ray Center (CXC) in the application packages CIAO, ChIPS and Sherpa. 

This research made use of Astropy\footnote{http://www.astropy.org}, a community-developed core Python package for Astronomy \citep{astropy13, astropy18}.

\section*{Data Availability}

The Gemini GMOS raw data used in this article is available to download at the Gemini archive website\footnote{https://archive.gemini.edu} (Program GN-2016A-Q-77, PI Couto). {\it HST-STIS} data can be obtained at the MAST archive\footnote{https://archive.stsci.edu} (Program 8881, PI Sambruna). The {\it Chandra} raw data used in this article is also available for download at the {\it Chandra} data archive website\footnote{https://cda.harvard.edu/chaser/} (OBSIDs: 11688, 11669, 12106, 12119, PI Siemiginowska). The reduced data underlying this article will be shared on reasonable request to the corresponding author.





\bibliographystyle{mnras}
\bibliography{refs}








\bsp	
\label{lastpage}
\end{document}